\documentclass[11pt]{article}
\usepackage{latexsym}
\usepackage{amssymb}
\usepackage{amsmath}
\usepackage[hypertex]{hyperref}
\usepackage{graphicx}
\usepackage{axodraw}

\newcommand{\Mpl}{M_{\rm Pl}}

\begin{document}

\pagestyle{empty}

\begin{flushright}
hep-ph/0509118 \\
\end{flushright}

\vspace{1.5cm}

\begin{center}
{\bf\LARGE The Abundance of Kaluza-Klein Dark Matter} \\[5mm]
{\bf\LARGE with Coannihilation}

\vspace*{1.5cm}
{\large Fiona Burnell$^{1}$ and Graham D. Kribs$^{2}$}
\\[5mm]

\mbox{$^1$\textit{Joseph Henry Laboratories, Princeton University,
Princeton, NJ 08544}} \\[1mm]

\mbox{$^2$\textit{Department of Physics and Institute of Theoretical 
Science,}} \\
\mbox{\textit{University of Oregon, Eugene, OR 97403}} \\

\vspace*{0.5cm}

\end{center}

\vspace*{1.0cm}

\begin{abstract}

In Universal Extra Dimension models, the lightest Kaluza-Klein (KK)
particle is generically the first KK excitation of the photon and 
can be stable, serving as particle dark matter.  We calculate the 
thermal relic abundance of the KK photon for a general mass
spectrum of KK excitations including full coannihilation 
effects with all (level one) KK excitations.  We find that including 
coannihilation can significantly change the relic abundance when
the coannihilating particles are within about 20\% of the mass
of the KK photon.  
Matching the relic abundance with cosmological data, 
we find the mass range of the KK photon is much wider than previously 
found, up to about 2 TeV if the masses of the strongly interacting 
level one KK particles are within five percent of the mass of 
the KK photon.  We also find cases where several coannihilation 
channels compete (constructively and destructively) with one another.  
The lower bound on the KK photon mass, about 540 GeV when just
right-handed KK leptons coannihilate with the KK photon, relaxes
upward by several hundred GeV when coannihilation with electroweak 
KK gauge bosons of the same mass is included.  

\end{abstract}

\newpage
\pagestyle{plain}
\section{Introduction}

One of the most important astrophysical challenges is to 
understand the nature and identity of dark matter.
Perhaps the most interesting candidate is a neutral, stable, 
weakly interacting particle arising from physics beyond the Standard Model.
This is consistent with a wide range of data, such as measurements
of the cosmic microwave background (CMB) radiation that imply
the abundance of dark matter is roughly a factor of 6 times the 
abundance of baryons \cite{WMAP}, as well as comparisons between
large scale structure simulations and galaxy survey observations 
that suggest the bulk of dark matter is cold \cite{structure}.

Universal Extra Dimension (UED) models \cite{ACD} (see also \cite{early}), 
in which all of the Standard Model fields propagate in extra
dimensions, 
provide an interesting example of Kaluza-Klein (KK) dark matter 
\cite{CMS,ST,KKother}.
This is because bulk interactions do not violate higher 
dimensional momentum conservation (KK number), and in these models
all of the couplings among the Standard Model particles arise from
bulk interactions.  To generate 4D chiral fermions, 
the extra compact dimension(s) must be  
modded out by an orbifold.  For five dimensions this is $S^1/Z_2$, 
while in six dimensions $T^2/Z_2$ is suitable and has other interesting 
properties \cite{ACD} including motivation for three generations
from anomaly cancellation \cite{DP} and the prevention of fast
proton decay \cite{ADPY}.  An orbifold does, however, lead to some
of the less appealing aspects of the model.  Brane-localized
terms can be added to both orbifold fixed points that violate KK number.
If these brane localized terms are symmetric under the exchange
of the two orbifold fixed points, then a remnant of KK number 
conservation remains, called KK parity.  All odd-level KK modes
are charged under this discrete symmetry thereby ensuring that the lightest 
level-one KK particle (LKP) does not decay.  The stability of the LKP 
suggests it could well be an interesting dark matter candidate.

We calculate the thermal relic abundance of dark matter
including coannihilation with all level-one KK particles.
This is an important calculation since there are many level-one 
KK particles that are near enough in mass to the LKP that 
they are expected to be important to determine the relic abundance.
The identity of the LKP depends on the mass
spectrum of the first KK level.  We assume it is
the lightest KK photon.  This has been widely advocated for both
its properties as dark matter and is favored by the contributions to the
masses of the level one KK particles from one-loop radiative corrections
\cite{CMS,KKother}.  
The spectrum of level-one KK particle masses is also not known.  However, 
certain finite and log-enhanced one-loop radiative corrections \cite{GGH} 
to the masses have been calculated in \cite{CMS}.  The size of
these corrections depend on two unknowns: the size of the log, i.e., the 
cutoff scale of the theory, and the size of the 
matching corrections evaluated at the cutoff scale.  
Generically one would expect the calculable log-enhanced 
corrections to dominate over the uncalculable matching corrections.
However, the log is not particularly large relative to the finite
matching corrections.
Indeed, should the
cutoff scale be close to the scale of the KK particle masses, 
uncalculable matching corrections would be of order the calculable 
log-enhanced corrections, and then one would like to know the
relic abundance for a general spectrum.  This is the calculation
we carry out in this paper.  

We present our results with three scenarios of level-one KK particle 
mass spectra.  First, we consider coannihilation
with all level-one KK particles that carry electroweak, but not color
charges.  Next, we consider coannihilation including colored
particles.  In both of these cases, we take all of the level one
KK particles (other than the LKP) to have the same mass
$m_{\gamma^{(1)}} (1 + \delta)$.
Finally, we consider coannihilation with all particles with a mass
spectrum derived from the loop corrections from \cite{CMS}.
We believe these three scenarios provide a good cross section 
of the effects of coannihilation.  We also provide formulae for
the (co)annihilation cross sections so that any other more complicated
spectra could easily be calculated based on our results.

The format of this paper is as follows.  
In Sec.~\ref{KKgeneral} we discuss KK dark matter including
past results and motivate the need for the calculations we carry out.
In Sec.~\ref{Freeze}, we
briefly describe the method used to calculate the relic abundance 
of the KK photon, and the effect of including coannihilating particles 
in this calculation.  In Sec.~\ref{Results}, we consider the effect of 
coannihilation on the relic abundance, and use this to obtain limits 
on the mass of the corresponding dark matter candidate, the KK photon.
We find mass ranges in three scenarios:  
(1) when electroweak KK excitations coannihilate, 
(2) when all KK excitations coannihilate, 
    taking all KK particles (except for the LKP) to have the same mass, and 
(3) when all KK excitations coanihilate assuming the mass spectrum is
    determined by the one-loop radiative corrections in \cite{CMS} 
    with zero matching corrections.
Appendices \ref{Lagrangian} and \ref{FeynmanRules}
reviews the Lagrangian and Feynman rules in UED\@.  
The diagrams and cross sections pertinent to these results are 
summarized in Appendices \ref{Diagrams} and \ref{CrossSect}.

\section{KK dark matter}
\label{KKgeneral}

The relic density of the KK photon was first calculated by
Servant and Tait \cite{ST}.  Using their results, one finds 
that its thermal relic abundance of the KK photon matches the 
WMAP observations for cold dark matter when the KK photon mass
is between about 550 to 850 GeV\@. 
The range of mass that they found depended on the relative importance 
of coannihilation with the level-one KK excitations of the right-handed
leptons.  This was the only coannihilation channel that was
considered in \cite{ST}.
Several groups have also examined the prospects for detecting
KK dark matter directly and indirectly \cite{KKother}.

Coannihilation is likely to occur through other level one KK excitations.
Examining a typical spectrum of KK excitations from the one-loop
radiative corrections discussed in \cite{CMS}, one finds that 
the KK excitations of the particles transforming under the
electroweak (but not color) groups tend to be within about 10\% 
of the mass of the KK photon.  This is near enough in mass that
coannihilation with these KK excitations is expected to affect
the the thermal relic abundance.  We emphasize that these 
radiative corrections, proportional to the log of the ratio
of the cutoff scale $\Lambda$ to the compactification scale $R^{-1}$,
are merely indicative
of shifts in the level one KK particle masses, since the finite 
matching corrections are unknown.  Indeed, if the cutoff scale
were not too large, even the KK excitations that transform under
SU(3)$_c$ could also play an important role in the calculation
of the thermal relic abundance.

What is the cutoff scale of UED models?  Generically, 
the cutoff scale is where the extra dimensional theory gets strong.
In the four dimensional effective theory this can be estimated by 
including KK excitations into loops, and one obtains typically 
$\Lambda R \sim 10$--$100$.
But there have also been recent explicit
scattering amplitude calculations that suggest the cutoff scale
may be much lower than previously assumed \cite{Chivukula:2003kq}.
In any case, it is clear that including the effects of coannihilation
for the entire spectrum will allow us to calculate the broadest range 
of KK photon
masses that lead to a thermal relic abundance consistent with
the cosmological measurements.
We will illustrate
the relative importance of different channels by showing the
results for specific choices of the level one KK spectrum.

The lower bound on the mass of the KK photon arises by 
calculating the effects of KK particles in loop corrections to 
precision electroweak data.  This was first done by \cite{ACD}
who found a lower bound of about $300$ GeV.  As we will see,
this bound is not saturated no matter what spectrum of 
level one KK excitations one takes, and so we do not need
to consider it further.

In general, the accuracy we desire requires only tree-level calculations.  
However a very recent paper did find an important effect on the relic
abundance of the KK photon \cite{KK2paper}.  The loop suppressed
operators on the orbifold fixed points lead to couplings between
pairs of level-one excitations and SM particles through
\emph{resonances} of level two (and higher) KK excitations.
The resonance behavior can roughly compensate for the loop suppression,
resulting in a large contribution to the total cross section.
Ref.~\cite{KK2paper} investigated the 
effect of the annihilation $\gamma^{(1)} \gamma^{(1)} \rightarrow t \bar{t}$ 
through an $s$ channel $h^{(2)}$, and found that including this process leads 
to a significant decrease in the relic abundance of $\gamma^{(1)}$.  
Including these loop-suppressed effects would make our calculations
prohibitively complicated, so in this paper we neglect them.
Instead, our motivation is orthogonal, namely to 
understand the implications of including coannihilation of all
level one KK excitations on the 
relic abundance.  Ultimately, to obtain the most accurate estimate of the
relic abundance of the KK photon would require taking into 
account both the resonance effect as well as the coannihilation 
effects that we report in this paper.

\section{The Relic Density of the KK photon}
\label{Freeze}

\subsection{Annihilation and freeze-out}

We now briefly review the standard method for calculating the 
cosmological relic density of a stable particle 
\cite{KolbTurner,GriestSeckel}.
The number density of a particle in the expanding universe is 
described by the Boltzmann equation
\begin{equation} 
\frac{d n_{\psi}}{dt} +3H n_{\psi} = - \left< \sigma_A v \right> 
\left[ (n_{\psi})^2-(n_{\psi}^{eq})^2 \right]
\label{Boltzmann}
\end{equation}
where $\left < \sigma_A v \right > $ is the thermal average of the total 
cross section for annihilation of the particle $\psi$ (assumed here to be 
equal to the total cross section for producing $\psi$) times the relative 
velocity of two particles, and $H$ is the Hubble expansion rate 
of the universe.  

We solve the Boltzmann equation to obtain the number density 
of a massive particle at late times.  
At early times ($T\gg m_\psi$, $n_\psi^{eq} \approx T^3$), $n_\psi^{eq}$ is 
large, and any deviation of 
$n$ from $n^{eq}$ rapidly goes to $0$.  At late times ($ T \ll m_\psi$, 
$n_\psi^{eq} \approx g(m_\psi T/2 \pi)^{3/2} e^{-m_\psi/T}$), $n^{eq}$ is 
tiny since the particles are now too massive to be thermally produced.
As $n_\psi$ decreases due to the expansion of the universe, the particles 
eventually become too dilute to annihilate
and freeze out at a constant density per 
comoving volume.  This freeze-out occurs roughly when the Hubble 
expansion rate overtakes the rate at which the 
particles annihilate, $\Gamma = \left< \sigma_A v \right > n_\psi\approx H$.

As we will verify later, KK dark matter freezes out at $T_f \ll m_{KK}$, 
so that at freeze out $n^{eq} = g \left (\frac{m^2}{2\pi x} \right) 
^{3/2} e^{-x}$, where  $x=\frac{m}{T}$, and $g$ is the 
number of degrees of freedom of the annihilating particle.  
In this case, (\ref{Boltzmann}) cannot be solved 
analytically.
Instead, the Boltzmann equation must be solved either numerically, or by 
means of a 
standard approximation which gives solutions consistent with numerical 
results to 
within $10\%$.  Here we shall use the latter approach.

To solve the Boltzmann equation approximately, we 
begin by making a simplifying change of coordinates. 
 If we assume the universe expands adiabatically, then 
$sa^3 =$~constant, where $s$ is the comoving entropy density 
given by $s = \frac{2 \pi^2}{45} g_* T^3$ and 
$g_*$ is the effective number of relativistic degrees of freedom.
In terms of the variables 
$Y = n_\psi/s$, $Y^{eq} = n^{eq} /s$, and $x=m/T$, 
(\ref{Boltzmann}) can be rewritten as
\begin{equation}
\frac{dY}{dx} = - \frac{\left <\sigma_A v \right> }{H x} s 
[Y^2-(Y^{eq})^2]
\end{equation}
where we assume that freeze-out occurs in a radiation dominated epoch, so 
that $\frac{ dx}{dt} =H x$.  Writing $\Delta =Y-Y^{eq}$, this becomes
\begin{eqnarray}
\Delta' + {Y^{eq}}' =& 
 -C \frac{\left< \sigma_A v \right >}{x^2} \Delta (2 Y^{eq}+ \Delta) 
\nonumber \\ 
=& -f(x) \Delta (2 Y^{eq}+ \Delta)
\end{eqnarray}
where $C=\sqrt{\frac{g_* \pi}{45}} m \Mpl$, and 
$f(x) = C \left< \sigma_A v \right > x^{-2}$.  

At early times, the particles track their equilibrium values, so that 
$\Delta \ll Y^{eq}$ and $\Delta' \ll {Y^{eq}}'$.  In this case
\begin{equation}
\Delta = -\frac{{Y^{eq}}'}{f(x) (2 Y^{eq} +\Delta)}
\end{equation}
To determine the freeze-out temperature, we use this expression for 
$\Delta$ to solve $\Delta = c Y^{eq}$ with $Y^{eq} = \frac{45 g}{\sqrt{32 
\pi^7} g_*} x^{3/2} e^{-x}$, 
where $c$ is a numerical factor whose optimal value is determined by 
comparison with numerical solutions of the Boltzmann equation 
($c=\sqrt{2}-1$).  
The freeze-out temperature can be deduced by solving the following 
equation numerically:
\begin{equation} 
x_f = \ln \left( c(c+2) \sqrt{\frac{45}{8}} \frac{g m \Mpl \langle \sigma_A v
\rangle}{2 \pi ^3 \sqrt{g_* x_f}} \right)
\label{GetT}
\end{equation}

At late times, $Y^{eq}$ and ${Y^{eq}}'$ are both small, and the equation 
reduces to 
\begin{equation}
\frac{\Delta ' }{\Delta^2} = -f(x)
\end{equation}
This equation is separable, and can thus be solved (with boundary 
conditions again chosen to give the best fit to numerical simulations): 
\begin{eqnarray} \label{GetOmega} 
Y_{\infty} \approx \Delta_{\infty} &= &\frac{1}{\int_{x_f}^{\infty} f(x) dx} 
\nonumber \\
\rho_{\psi} = m_{\psi} n_{\psi} |_{t=\infty}& =& m_{\psi} s_0 
Y_{\infty} \nonumber \\
\Omega_{\psi} = \frac{\rho_{\psi}}{\rho_c}& = &\frac{s_0 
\sqrt{45}}{\sqrt{\pi g_*} \Mpl \rho_c \int_{x_f}^{\infty} \left< 
\sigma_A v \right > x^{-2} dx }  \nonumber \\
&\approx& \frac{1.04 \times 10^9 }{\int_{x_f}^{\infty} \left< 
\sigma_A v \right > x^{-2} dx }
\end{eqnarray}
Note that because of the factor of $s_0$ appearing in the numerator, this 
solution assumes that the universe expands adiabatically between 
freeze-out and today.  This is, of course, not 
quite true, as some standard model particles will subsequently fall out of 
equilibrium, slightly increasing the entropy per comoving volume.  In 
practice, however, this contribution from known standard model sources is 
very small, and hence may safely be neglected.  

\subsection{Coannihilation}

Different species of interacting particles that have masses 
nearly degenerate with the KK photon will fall out of equilibrium at 
the same time and with roughly similar density.
Interactions of the form $\psi_1 X \rightleftharpoons 
\psi_2 X'$ (where $\psi_1$ and $\psi_2$ are the KK particles, and 
$X,X'$ are SM particles), converting the relic into other particles of 
similar mass, occur rapidly.  In fact, such processes are much more
efficient than the relevant annihilation processes, so that the abundances 
of the all particles are effectively correlated during freeze-out.  
The slightly heavier particle species will eventually 
all decay to the lightest stable particle (in our case, the KK photon)
and thus contribute to the relic density.    This situation 
is called coannihilation.

As shown in \cite{GriestSeckel}, to compute the relic density and freeze-out 
temperature with 
coannihilation, it suffices to substitute an effective cross section 
$\sigma_{\rm eff}$ for $\sigma$ in Eqs.~(\ref{GetOmega}),(\ref{GetT}) 
above, where
\begin{eqnarray}
\sigma_{\rm eff}=\sum_{i,j} \sigma_{ij} \frac{g_i g_j}{g_{\rm eff}^2} 
(1+\Delta_i)^{3/2}(1+\Delta_j)^{3/2} e^{-x(\Delta_i+\Delta_j)} \nonumber \\
g_{\rm eff}= \sum_i g_i(1+\Delta_i)^{3/2} e^{-\Delta_i x} 
\end{eqnarray}
The relic abundance depends on a weighted average of the annihilation 
and coannihilation cross sections of all relevant particles. 
Thus if the coannihilating particles interact more strongly compared 
with the annihilating particles (while also being close enough in mass)
they can 
increase the effective cross section and thereby decrease the relic 
density.  Conversely, if they are more weakly interacting, then 
the coannihilation effects decrease the effective cross section, 
yielding a larger final value for $\Omega_{\psi}$.

A simple intuitive picture of the effects of including coannihilation 
comes from noting that
\begin{equation}
\rho \sim \left(\int_{x_f}^\infty \left <\sigma_A v \right > x^{-2} 
\right)^{-1} .
\end{equation}
The integral over $x$ will of course result in a different weight for 
$s$ and 
$p$ wave annihilation, nevertheless it is easiest to simply ignore
this difference and approximate
$\rho_{\psi} \sim (\left < \sigma v\right> )^{-1}$.  
A larger effective cross section results in more annihilation 
during freeze-out, and thus a smaller relic density.  

Furthermore, it is easy to see that the effective cross section  
$\sigma_{\rm eff}$ is a weighted average 
of the relevant annihilation and coannihilation cross sections by 
considering 
the case where all but the lightest (first) coannihilating particles 
have the same mass $m_j= m_0(1+\delta)$, $j\neq 1$, and all particles 
have the same number of degrees of freedom.
Then
\begin{eqnarray}
A &=& \frac{\sum\limits_{ij} 
      (1+\delta)^{3 (a_i + a_j)/2} e^{-(a_i + a_j) \delta x} 
      \frac{\sigma_{ij}}{\sigma_{11}}}{\sum\limits_i(1+\delta)^{3 a_i/2} 
      e^{-a_i x \delta}} \nonumber \\
      \delta \rightarrow 0 &\Rightarrow & \frac{1}{N^2} \sum_{ij} 
      \frac{\sigma_{ij}}{\sigma_{11}} 
\end{eqnarray}
where $i,j$ run over all included particles, 
$a_{i,j} = 0 (1)$ for the LKP (all other KK particles), 
and $\sigma_{ij}$ is the total (co)annihilation cross section for 
the process $ \psi_i \psi_j \rightarrow$ Standard Model particles.

This makes it clear that in the limit of a small mass difference and 
equal numbers of degrees of freedom, 
$\sigma_{\rm eff}$ is precisely the average of all relevant 
cross sections while
the relic abundance depends only on whether this average is larger 
or smaller than $\sigma_{11}$.  If all 
$\sigma_{ij}$ are equal, then including coannihilation does not alter the 
relic density; if the weighted cross sections of the 
coannihilating particles are on average larger (smaller) than that 
of the original particle, then the relic density will be smaller  (larger) 
when coannihilation is included. 
This holds irrespective of the value of 
$\delta$.  However, since $\delta$ enters exponentially 
into the weights of the coannihilation
cross sections, the effects of coannihilation are rapidly suppressed as the 
relative mass difference $\delta$ increases.  Typically for weakly
interacting particle dark matter,
$x_f \approx 20$ to $30$, and for $\delta > 0.1$ the effects of 
coannihilation are generally negligible.  However, for very large 
coannihilating cross sections, fractional mass differences up to 
$\delta =0.2$ can affect the relic density \cite{GriestSeckel}.  

\section{Results} 
\label{Results}

Now we are ready to consider the effect of coannihilation on the 
relic abundance of the the KK photon.  Each
coannihilation scenario considered requires several total annihilation 
cross sections; the relevant Feynman diagrams, together with tables of 
the formulae of the relevant cross sections, are presented in Appendices 
\ref{Diagrams} and \ref{CrossSect}.  

Certain approximations were made in obtaining our results.
Cross sections were computed ignoring all 
terms of order $\frac{m_{SM}}{m_{KK}}$ where $m_{SM}$ is any standard 
model mass.  Consequently, we ignored all Yukawa couplings which are 
proportional to the corresponding fermion mass.  
This approximation is well justified (in that including such terms should 
alter the relic abundance by less than $10 \%$) for all particles 
except possibly the top quark, whose mass $m_t \approx 175$ GeV is 
close to one half of the precision electroweak lower bound on $R^{-1}$.
In particular, as shown in \cite{KK2paper}, 
the top Yukawa coupling can alter the cross section 
significantly, as it leads to nearly resonant $s$-channel diagrams.  
Aside from this effect, our results are not expected to be sensitive
to the top mass since, as we will see, the masses of the KK photon 
necessary to explain the observed relic dark matter abundance are 
well above the mass of the top quark.
It addition, we neglected the mixing between $B^{(1)}$ and
$W^{3(1)}$, which are expected to be rather small already for
the first KK level (see Appendix \ref{KWeinS}).
Hence, we take $\gamma^{(1)} \equiv B^{(1)}$, $Z^{(1)} \equiv W_3^{(1)}$ 
throughout.

When examining the relic density as a function of mass, a further 
simplification is obtained by ignoring the mass-dependence of $x_f$.  
The value of $x_f$ depends very weakly on the mass.  Indeed, 
from (\ref{GetT}) we expect this dependence to be approximately 
logarithmic.  Typically, over the mass range of $m_{KK} = 0.2$ to $2$ TeV, 
$x_f$ varies by about $2$ GeV/degree, or less than 10\%.\footnote{The 
dependence of $x_f$ on the relative masses of the coannihilating 
particles, however, has a 
somewhat larger effect, and have been taken into account.}  This variation 
has a small effect on the relic density.  This also shows that
the KK dark matter is cold: for all cases considered here, 
we find $23 < x_f < 28$, so that the particles are well approximated 
as non-relativistic.

We use couplings and astrophysical parameters that can be obtained
from \cite{PDG}.  In particular, we take the electroweak and
strong couplings evaluated at $M_Z$ (ignoring the renormalization 
group running of the couplings).  In practice, the freeze-out temperature 
is between about $25$ and $100$ GeV, depending on which particles 
coannihilate.  The dark matter relic density is taken to be the 
$1\sigma$ value from WMAP, $\Omega_{DM} = 0.113 \pm 0.01$ \cite{WMAP}.

\subsection{Coannihilation with simplified level-one KK spectrum}

To investigate the effects of KK particles coannihilating
with the KK photon, we chose two different mass 
spectra for the coannihilating particles.  First, particles in the first
KK level are divided into two groups:  those with masses 
approximately $m_{\gamma^{(1)}} (1+ \delta)$, with $\delta$ small, 
and those that are too heavy to play a role in coannihilation.  While 
this choice of mass spectrum gives a good understanding of the range of 
possible outcomes of including coannihilating particles, it is also 
somewhat arbitrary.  Thus, we also investigate a mass spectrum based on 
that derived in \cite{CMS} in the next subsection.

The first check on our results was to consider coannihilation of the 
KK photon with just right-handed leptons, where our calculations and 
numerical results agree with \cite{ST}.  This provides a non-trivial 
check on our procedure for calculating the relic density in the
Universal Extra Dimensions model.

\begin{figure}[htp] 
\centering
\includegraphics[totalheight=.6\textheight]{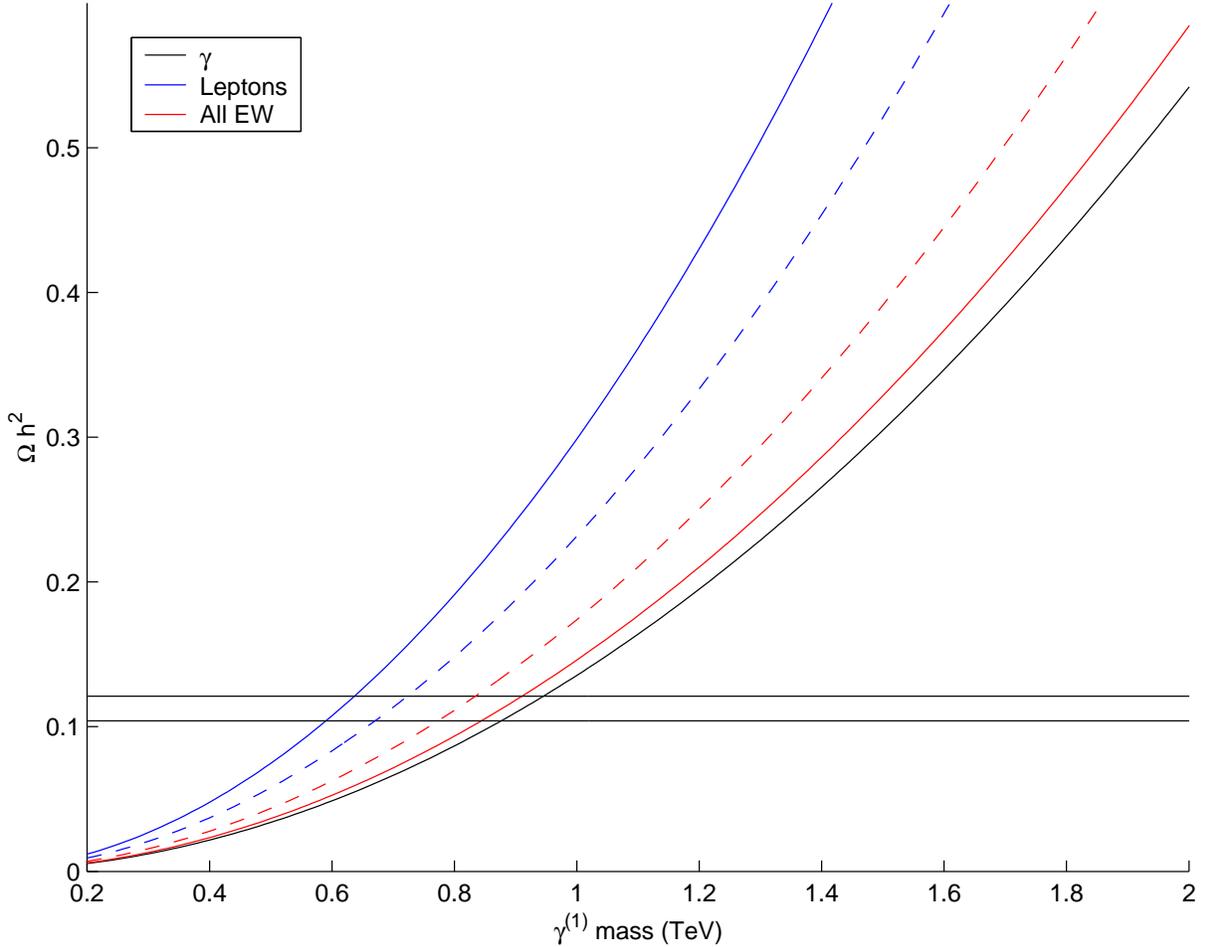}
\caption{Relic abundance of KK dark matter as a function of mass 
after including no coannihilation (black; right-most solid line),
coannihilation of $B^{(1)}$ with all leptons 
(blue; left-most pair of solid and dashed lines), 
and all electroweak particles (red; middle pair of solid and dashed lines), 
assuming in each case that all coannihilating particles have the same 
mass $m_{\gamma^{(1)}}(1+\delta)$.  The solid (dashed) lines 
show the values for $\delta = 0.01$ $(0.05)$ for the cases with
coannihilation.  Notice that for the case including KK electroweak 
gauge bosons, the abundance as a function of mass is smaller
for the smaller mass splitting $\delta = 0.01$, 
unlike the case with just KK leptons, 
since the effects of coannihilation of these two sets of KK particles 
somewhat compensate for each other in the relic abundance calculation.}
\label{AllEW} 
\end{figure}

Figure \ref{AllEW} shows the effect of including coannihilation of 
the KK photon with leptons (blue), and all electroweak particles
including leptons, scalars, and electroweak gauge bosons (red).
Here we assume in each case that particles included in the coannihilation 
have the same mass $m_\gamma (1+\delta)$.
The graph shows that coannihilation with leptons, and to a lesser extent 
with $W$ and $Z$ bosons, tends to increase the relic density for a given 
mass of the KK photon.  

\begin{figure}[htp] 
\centering
\includegraphics[totalheight=.6\textheight]{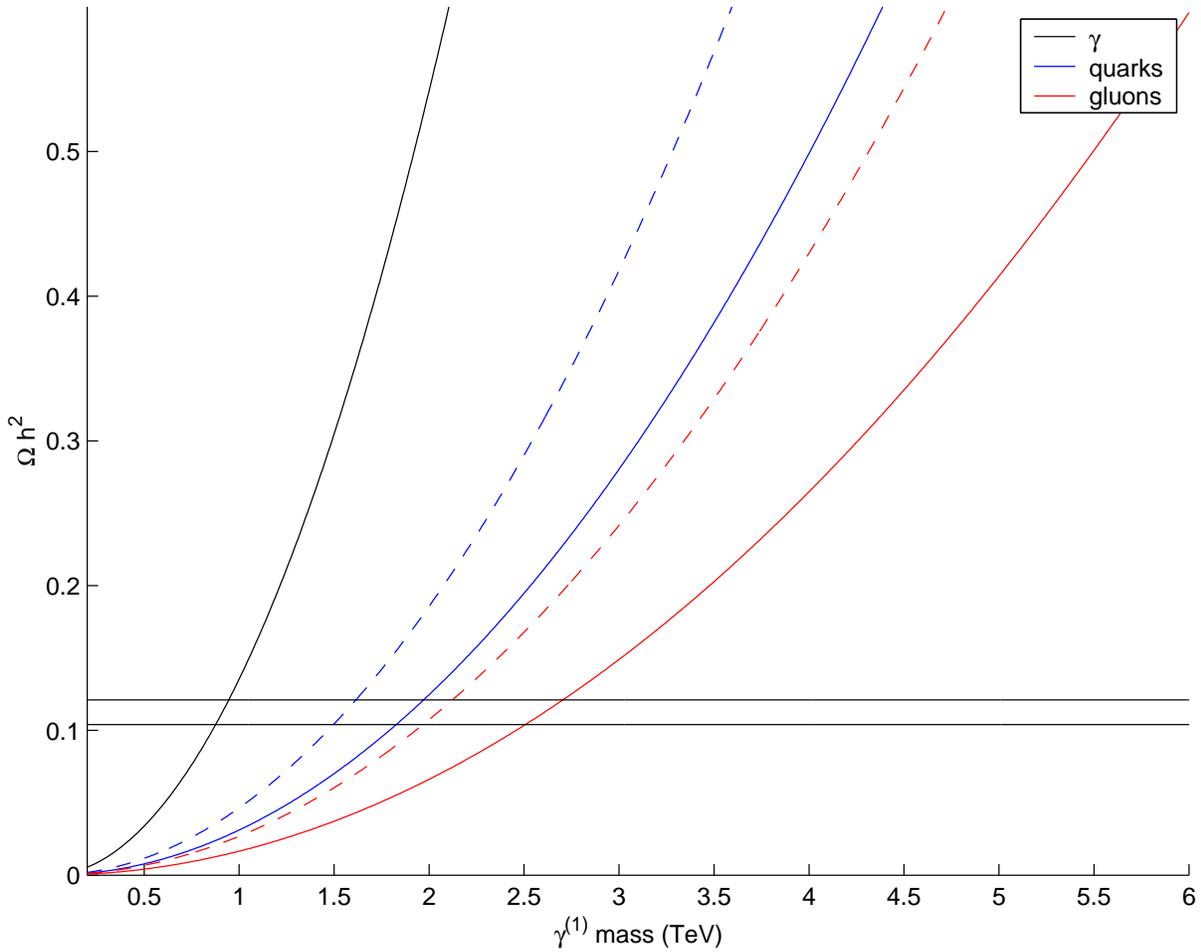}
\caption{Relic abundance of KK dark matter as a function of mass 
after including coannihilation of $B^{(1)}$ with  
 all non-strongly interacting particles {\it and} quarks (blue), and 
all level one particles (red)
 assuming in each case that all coannihilating particles 
have the same mass $m_{\gamma^{(1)}} (1 + \delta)$.  The solid 
and dashed lines show the values for $\delta = 0.01$, 
and $\delta = 0.05$.}
\label{AllEW2} 
\end{figure}

Figure \ref{AllEW2} shows the effect of including coannihilation of 
the KK photon with all level one KK electroweak particles and level one
KK quarks (blue), and all particles at KK level one including
the KK gluon (red).
Again, we assume in each case that all particles included in the 
coannihilation have the same mass $m_\gamma (1+\delta)$.
Clearly, including coannihilation with strongly interacting KK particles
decreases the relic abundance for a fixed KK photon mass.
The extent of this decrease is one of the most important results
of this paper.  In particular, if the level one KK particles 
are highly degenerate to within a few to perhaps 10\%, the KK photon 
mass consistent with the thermal relic abundance that matches the WMAP data
could be several TeV.  

\begin{figure}[htp] 
\centering
\includegraphics[totalheight=.5\textheight]{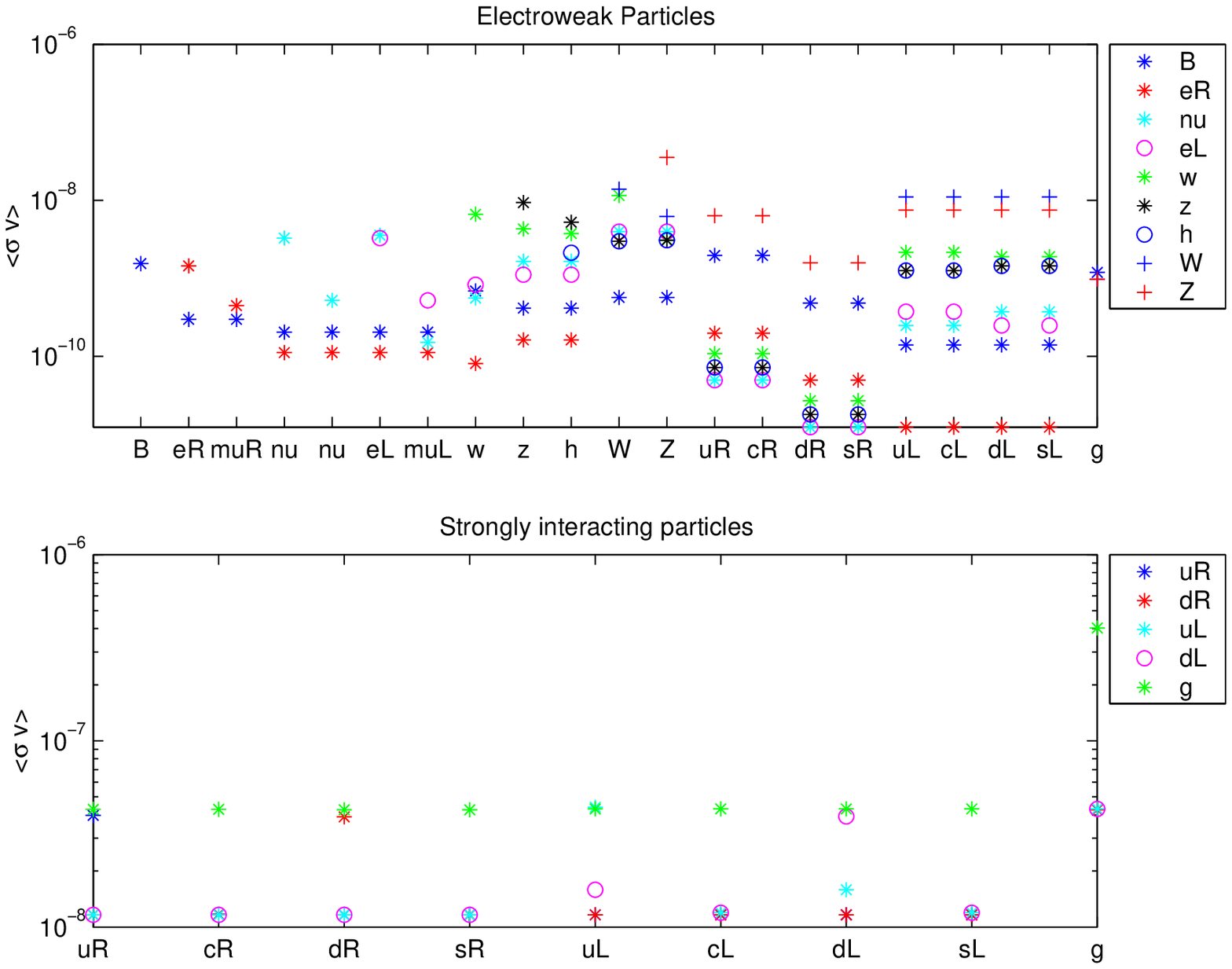}
\caption{Magnitudes of all cross sections, using  $m_{KK}=1$ TeV, 
$x_f=25$, and 
all couplings taken at the scale $M_Z$.  Here each color 
represents a particular particle species (see legend), and the x-axis 
indicates with which particle the (co)annihilation is occurring.  For 
example, a blue dot above the text $e_L$ is the coannihilation 
cross section of $B^{(1)} \equiv \gamma^{(1)}$ with $e_L$. Only 2 families of 
fermions are shown, as this is sufficient to show the difference between 
annihilation between members of the same family, and coannihilation between 
members of different families.  The 
annihilation cross sections for fermions, 
$W$ and its scalar counterpart $w$ have 
been weighted by a factor of $1/2$ to account for the two helicities.
In general the coannihilation 
cross sections are smaller than the annihilation ones, (and more numerous) 
so that usually adding more coannihilating particles decreases the 
effective cross section.
However, as the SU(2) coupling is stronger than the U(1) coupling, 
and as it also opens more channels, when  scalars and gauge 
bosons, whose annihilation cross sections are quite large, are added,  the 
average cross section increases.}
\label{EWCrossSections} 
\end{figure}

These results can be more readily understood by examining the 
relative magnitudes of the cross sections in question, shown in
in Fig.~\ref{EWCrossSections}.  Notice first that
the coannihilation cross section of any non-colored particle with 
$B^{(1)}$ is 
smaller than the annihilation cross section of $B^{(1)}$ with itself.  
Thus 
coannihilation with particles whose self-annihilation cross sections 
are close to those of $B^{(1)}$ tends to decrease the effective 
cross section, and hence increase the relic abundance for a given 
$B^{(1)}$ mass.  This is the case for leptons and scalars.  The 
level one KK electroweak gauge bosons $W^{(1)}$ and $Z^{(1)}$ 
and all strongly interacting particles have sufficiently large 
self-annihilation cross sections that they cause an increase in the 
effective cross section, and decrease the relic abundance.  For
$W^{(1)}$ and $Z^{(1)}$ this effect is relatively small, 
and their inclusion does not completely counter-balance the 
reduction in $\sigma_{eff}$ resulting from 
including leptons and scalars.  In the case of the strongly interacting 
particles, the relevant cross sections are approximately an order 
of magnitude larger than those of the non-strongly interacting particles, 
and the resulting reduction in the relic density is quite dramatic 
at small $\delta$.

\subsection{Coannihilation with the Cheng, Matchev, Schmaltz spectrum}

In this section we consider the spectrum that results if one
takes the one-loop radiative corrections to the KK masses 
from \cite{CMS}.  There are several assumptions built into
this spectrum.  One is that the matching contributions to the
brane-localized kinetic terms are assumed to be zero when
evaluated at the cutoff scale.  Furthermore, the
radiatively generated
terms are log enhanced by a log of the ratio of the cutoff scale 
$\Lambda$ to the mass of the KK excitation.  This ratio is also not known,
and may be much smaller than previously estimated \cite{Chivukula:2003kq}.
We therefore consider a wide range of $\Lambda R$ to illustrate
the effects of coannihilation.

First, let us summarize the results of \cite{CMS} that are used 
for the masses of the KK excitations in this Section:
\begin{eqnarray} \label{MassCorr}
\delta(m^2_{B^{(n)}})& =& \frac{ g'^2}{16 \pi^2 R^2} \left ( 
\frac{-39}{2} \frac{\zeta(3)}{\pi^2} -\frac{n^2}{3} \ln \, 
\Lambda R \right) \nonumber \\
\delta(m^2_{W^{(n)}}) &=& \frac{ g^2}{16 \pi^2 R^2} \left ( 
\frac{-5}{2} \frac{\zeta(3)}{\pi^2} + 15 n^2 \ln \,
\Lambda R \right )\nonumber \\
\delta(m^2_{g^{(n)}}) &=& \frac{ g_s^2}{16 \pi^2 R^2} \left ( 
\frac{-3}{2} \frac{\zeta(3)}{\pi^2} + 23 n^2 \ln \,
\Lambda R \right )\nonumber \\
\delta(m_{Q^{(n)}})&=& \frac{n}{16 \pi^2 R} \left ( 6 g_s^2+ \frac{27}{8} 
g^2 + \frac{1}{8} g'^2 \right) \ln \, \Lambda R \nonumber \\
\delta(m_{u^{(n)}})&=& \frac{n}{16 \pi^2 R} \left ( 6 g_s^2+  
 2 g'^2 \right) \ln \, \Lambda R \nonumber \\
\delta(m_{d^{(n)}})&=& \frac{n}{16 \pi^2 R} \left ( 6 g_s^2+  \frac{1}{2} 
g'^2 \right) \ln \, \Lambda R \nonumber \\
\delta(m_{L^{(n)}})&= &\frac{n}{16 \pi^2 R} \left ( \frac{27}{8} 
g^2 + \frac{9}{8} g'^2 \right) \ln \, \Lambda R \nonumber \\
\delta(m_{e^{(n)}})&= &\frac{n}{16 \pi^2 R}\frac{9}{2} g'^2 \ln \, \Lambda R 
\end{eqnarray}
All of the non-colored KK excitation masses are within about 10\% 
of $m_{\gamma^{(1)}}$ up to moderately high values of the cutoff scale 
($\Lambda R \approx 30$), 
and hence are almost certainly relevant for coannihilation.  
Conversely, for $\Lambda R > 5$, the masses of the strongly interacting 
particles
are more than $10 \%$ greater than $m_{\gamma^{(1)}}$, and thus are 
less likely to be important for coannihilation.  However, as emphasized
by \cite{GriestSeckel}, particles with sufficiently large cross sections 
may be relevant to coannihilation even for mass differences of up to 
of order $20 \%$.

We further simplify the mass spectrum given by (\ref{MassCorr}), 
while ignoring the SM top quark mass, by dividing the KK particles 
into five mass classes: 
the KK photon; 
the KK leptons and scalars; 
the KK $W$ and $Z$ gauge bosons; 
the KK quarks; and 
the KK gluon.
(In cases with more than one particle of slightly differing mass,
we take the average mass of all of the particles in that class.)

\begin{figure}[htp] 
\centering
\includegraphics[totalheight=.5\textheight]{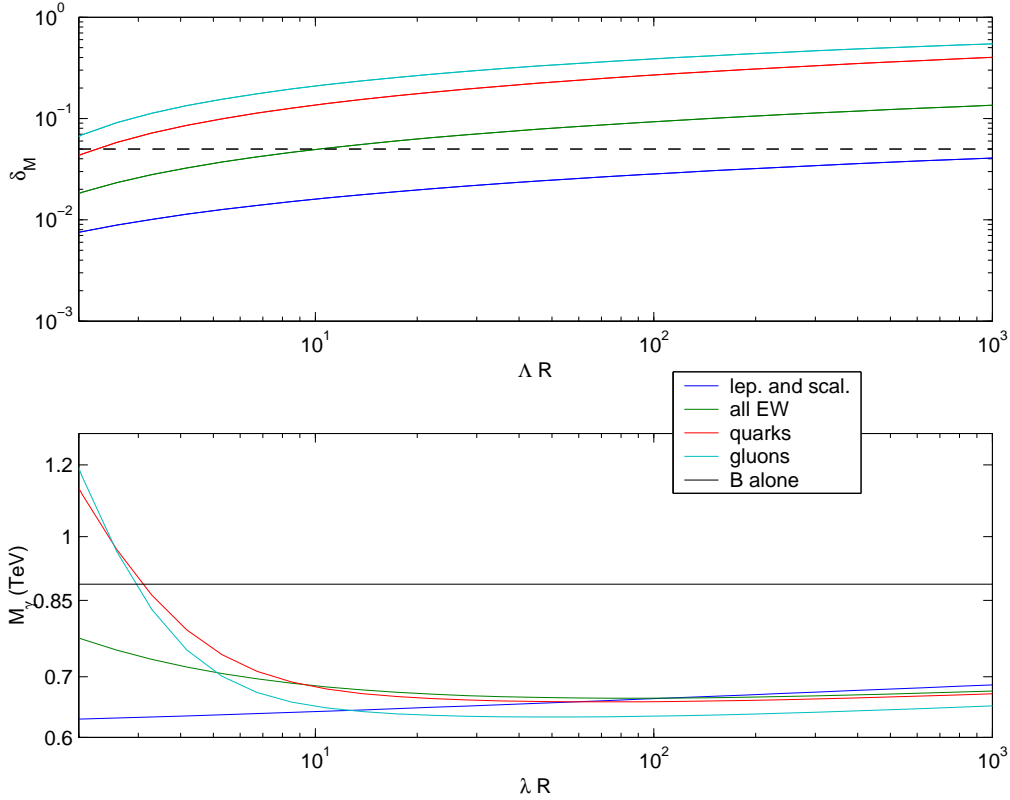}
\caption{The calculated $\gamma^{(1)}$ mass yielding a thermal
relic abundance consistent with WMAP observations is shown
in the bottom graph using as a function of $\Lambda R$.
This result was calculated using a simplified spectrum 
of the first KK level consisting of four sub-levels that
vary logarithmically with $\Lambda R$, shown in the top graph.
The ratios are from top to bottom:  the KK gluon, the KK quarks,
the KK electroweak gauge bosons, and the KK leptons and scalars,
divided by the KK photon mass.
The black dashed line corresponds to $\delta = 0.05$ for reference.
In the bottom graph, including increasing numbers of particles at
the first KK level are shown with the four lines labeled in the legend.
The solid straight black line (the line at the top on the far right) 
shows the the case without coannihilation for reference.
All level one KK particles except for the leptons and scalars 
become rapidly irrelevant to the relic abundance calculation once
$\Lambda R \gtrsim 100$.}
\label{ADSMass} 
\end{figure}

Figure \ref{ADSMass} shows the $\gamma^{(1)}$ mass that results in a
thermal relic abundance consistent with cosmological data, as a 
function of $\Lambda R$, given a five class mass spectrum.  
The upper plot also shows the mass gap $\delta$ for each of the classes.
Because $g_s$ is considerably larger than $g,g'$, the masses of the 
strongly interacting particles increase rather quickly as the cutoff 
increases.  For $\Lambda R \geq 2$, nearly all strongly 
interacting particles have $\delta > 0.05$.  
Masses of the leptons and scalars, however, vary rather slowly 
with $\Lambda R$, and remain within $10 \%$ of the 
$\gamma^{(1)}$ mass out to values $\Lambda R \sim 10^{4}$.  
Thus as the cutoff is increased, the gluons, quarks, and to a lesser 
extent the $W$ and $Z$ bosons, rapidly become too heavy to play a role 
in coannihilation, and the relic density is determined by the 
effective cross section of $\gamma^{(1)}$ and the KK leptons and scalars.
Thus the mass spectrum, Eq.~(\ref{MassCorr}), favors a somewhat lower
value of $m_{\gamma^{(1)}} \approx 700$ GeV to be consistent with the 
measurements of the dark matter abundance in the Universe.

\section{Conclusions}

We have calculated the thermal relic abundance of the KK photon
in the five dimensional UED model for a generalized mass spectrum
of level one KK particles.  We find that the lowest KK photon mass 
which could possibly account for the observed dark matter relic 
abundance is $540$ GeV, resulting from including coannihilation with 
three generations $e_R$ and neutrinos all very nearly degenerate with 
$\gamma^{(1)}$ ($\delta =0.01$).  This is consistent with the result
found in Ref.~\cite{ST} updated to reflect the WMAP measurements.
Since the radiative mass corrections should be the same 
for $e_L$ and $\nu$, a more realistic estimate, given by including all 
KK leptons, is $570$ GeV.  This is significantly lower than the lower 
mass bound of $860$ GeV given by including $\gamma^{(1)}$ alone.
Including level one KK quarks and the KK gluon with masses within
about 10\% of the mass of the KK photon significantly increases
the total effective annihilation cross section.  This implies the
KK photon mass leading to a thermal relic abundance consistent 
with WMAP observations is much larger, up to several TeV, 
see Figs.~\ref{AllEW2} and \ref{ADSMass} for more precise numerical ranges.
On face value, such a small separation between the KK photon and
the strongly interacting level one KK particles is not expected 
from the radiative corrections to the masses of the first KK level 
computed in \cite{CMS}.  However, if the cutoff scale is not much
larger than the KK photon mass itself, and thus matching corrections 
are comparable in size while opposite in sign to compensate, 
the level one KK spectrum could be much more degenerate.
These results show that the range of the KK photon mass is
much wider if indeed the mass spectrum is more degenerate than 
previously thought.  Given measurements of the level one KK particle 
masses, the calculations presented here could be used to find
the total effective cross section to verify if the KK photon
does (or does not) make up the dark matter density needed
to be consistent with WMAP observations.

{\bf Note added:} During the completion of this work we became aware of
an analogous calculation done independently by another 
group~\cite{KongMatchev}.
We have compared extensively the formulas for the annihilation 
cross-sections involving KK quarks, KK leptons and KK gauge bosons, 
in the limit of degenerate KK masses, as listed in the Appendix. 
In all considered cases we found perfect agreement.


\section*{Acknowledgments}

We are grateful to K.C.~Kong and K.~Matchev for extensively 
comparing their results with ours.
We thank G.~Servant and T.~Tait for discussions.
GDK thanks the Aspen Center for Physics for hospitality
where part of this work was completed.
This work was supported in part by NSERC and by DOE under contracts
DE-FG02-96ER40969 and DE-FG02-90ER40542.


\appendix
\section{Universal Extra Dimensions} 
\label{Lagrangian}

Here we provide a brief overview of the Universal Extra Dimension
model in five dimensions (for a review, see \cite{KribsTASI}).
In subsequent Appendices we list the Feynman diagrams for 
(co)annihilation and the cross section results we obtained.
The particle content of the UED model is shown in Table \ref{Tab1}.
The $Z_2$ orbifold projects out one helicity of the fermion zero modes, 
as well as the zero mode of the $5^{th}$ component of the gauge field.
We begin with a brief survey of the mass eigenstates of the theory; the 
Feynman rules are presented in Appendix \ref{FeynmanRules}.

\begin{table}[h] 
\begin{center}
\begin{tabular} {|c|c|c|}
\hline 
$5D$ field & $4D$ Even Fields & $4D$ Odd Fields \\
\hline 
$A_M$ & $A_\mu$ & $A_5$ \\
$\left(L, \tilde{L} \right)$& $L=\left (\nu, e_L \right )$ & 
$\tilde{L}=\left 
(\nu_R, e_R \right)$ \\
$\left(Q, \tilde{Q} \right)$& $Q =\left (u_L,  d_L \right)$ &  
$\tilde{Q}=\left (u_R,  d_R \right)$\\
$\left( e, \tilde{e} \right)$& $ e=e_{R}$ & $\tilde{e}=e_L$ \\  
$\left( u, \tilde{u} \right)$ & $ u= u_{R}$& $\tilde{u}= u_{L}$  \\
$\left( d, \tilde{d} \right)$ & $ d= d_{R}$& $ \tilde{d}=d_{L}$  \\
$\phi$ & $\phi$ & \\
\hline 
\end{tabular}
\end{center}
\caption{$5D$ fields and their behavior under the orbifold projection
$y \rightarrow -y$.}
\label{Tab1}
\end{table}

\subsection{Fermions}

The kinetic terms for fermions have the form:
\begin{equation}
L_{kinetic} = \overline{\psi}(x,y)(\gamma^\mu \partial_\mu +g_{5,5}\gamma_5 
\partial_5) \psi(x,y) \end{equation}
where we have ignored SM mass terms.  Upon doing the usual 
Fourier expansion for, for example, the first generation of 
(left-handed) quarks, and integrating over $y$, we obtain:
\begin{eqnarray}
L_{kinetic} = &\overline{(u,d)_L}\gamma^\mu \partial_\mu (u,d)_L  
+\sum\limits_{j=1}^{\infty} \Big[ \overline{P_L Q_{1L}^{(j)}}
\left( \gamma^\mu \partial_\mu P_L Q_{1L}^{(j)} 
-\gamma^5 \frac{j}{R} P_R Q_{1R}^{(j)} \right) \nonumber \\
&+\overline{P_R Q_{1R}^{(j)}}\left( \gamma^\mu \partial_\mu P_R Q_{1R}^{(j)} 
+\gamma^5 \frac{j}{R} P_L Q_{1L}^{(j)} \right)  \Big] \; .
\end{eqnarray}
(We use lower case letters to denote SM particles, and upper case
for their KK excitations.)  
To obtain the full fermion masses, the usual mass terms arising from 
the Yukawa couplings are added (then diagonalizing 
the resulting mass matrices).

\subsection{Gauge Bosons}

After compactification, the kinetic terms in the gauge boson 
Lagrangian can be expressed as
\begin{equation}
L_{kinetic}= -\frac{1}{4}\left ( F_{\mu\nu} F^{\mu \nu} + 
2 (\partial_5 A_\mu -\partial_\mu a)^2 \right )
\end{equation}
where $A_\mu$ is the 4D gauge field, and (to make the 
analogy with the Higgs mechanism more apparent) $a=A_5$, a 4D 
scalar.  Integrating over the fifth dimension coordinate $y$ 
causes all cross-terms between modes of different KK-number to 
cancel, leaving
\begin{equation}
L_{kinetic}= -\frac{1}{4} \sum_{j=0}^{\infty} 
\left (\partial_{\mu} A^{(j)}_{\nu} - \partial_{\nu}A_{\mu}^{(j)}
\right )^2 + 
\frac{1}{2} \sum_{j=1}^{\infty} \left (\frac{j}{R} 
A_{\mu}^{(j)} - \partial_\mu a^{(j)} \right)^2  \; .
\end{equation}
Expanding the second term, and collecting modes of different KK number, 
we obtain:
\begin{eqnarray} 
L_{kinetic} &=& -\frac{1}{4}
\left(\partial_{\mu} A^{(0)}_{\nu} - \partial_{\nu}A^{(0)}_{\mu} \right)^2 + 
\sum\limits_{j=1}^{\infty} \bigg[ -\frac{1}{4}
\left( \partial_{\mu} A^{(j)}_{\nu} - \partial_{\nu}A^{(j)}_{\mu}\right)^2  
  \nonumber \\
& &{} + \frac{1}{2} \left (\frac{j}{R}\right)^2 
A_{\mu}^{(j)}A^{\mu (j)} - \left( \frac{j}{R} \right) \partial_\mu 
a^{(j)}A^{\mu (j)} +\frac{1}{2} \partial_\mu a^{(j)}
 \partial^{\mu} a^{(j)} \bigg]
\label{GKin}
\end{eqnarray}
This is precisely the Lagrangian for a tower of 4D gauge fields with 
masses $M_j = \frac{j}{R}$ generated by a spontaneously broken symmetry.
The scalar fields $a^{(j)}$ are eaten by the gauge field 
$A^{(j)}_\mu$ in the usual Higgs mechanism to give them mass.

\subsection{Electroweak KK Gauge Boson Mass Eigenstates} 
\label{KWeinS}

There is a tower of KK gauge bosons for each of the SM gauge symmetries.
The electroweak gauge boson KK tower, however, is different from
the SM zero modes in the admixture of $W^{3(j)}$ and $B^{(j)}$.
This is because the KK $SU(2)$ and $U(1)$ gauge bosons receive different 
loop corrections to their masses.  The actual KK 
mass eigenstates can be found by diagonalizing the matrix
\begin{equation}
\left( \begin{array}{cc} 
  \frac{n^2}{R^2} + \delta_1 +\frac{1}{4} g^2 v^2 & \frac{1}{4} g g' v^2 \\
  \frac{1}{4} g g' v^2 & \frac{n^2}{R^2} + \delta_2 + \frac{1}{4} g^{'2} v^2 
\end {array} \right) 
\end{equation}
where $\delta_1= \delta(m_{W_3^{(n)}})$, $\delta_2= \delta(m_{B^{(n)}})$
are the radiative corrections to the $n^{\rm th}$ KK level electroweak
gauge bosons (precise expressions can be found in \cite{CMS}).  
We can re-express this as:
\begin{equation}
\left( \begin{array}{cc} 
  \frac{n^2}{R^2} + \Delta_M &             0              \\
             0               & \frac{n^2}{R^2} + \Delta_M  
\end {array} \right) + \left( \begin{array}{cc} 
  \delta +\frac{1}{4} g^2 v^2 & \frac{1}{4} g g'v^2 \\
  \frac{1}{4} g g' v^2        & -\delta + \frac{1}{4} g^{'2} v^2 
\end{array} \right)
\end{equation}
where $\Delta_M = \frac{\delta_1 +\delta_2}{2}$, and $\delta = 
\frac{\delta_1-\delta_2}{2}$.   
The first contribution to the masses is diagonal in the $W^{(3)}$-$B$ basis,  
and does not affect mixing.  The mass eigenstates result from diagonalizing 
the second matrix; thus the Weinberg angle at each KK level is determined 
by the relative sizes of $\delta$ and the electroweak masses 
$\frac{1}{4} g_i g_j v^2$.  Since $\delta$ is proportional to $\frac{1}{R^2}$,
if $m_{KK} \gg M_{ew}$, the KK mass matrix is almost diagonal in 
the $W^{(3)}$-$B$ basis.  Since $m_{KK}$ is at least 300 GeV \cite{ACD},
the KK Weinberg angle is much smaller than its SM counterpart.
In fact, for $R^{-1} \geq 600$ GeV (which is roughly the lower bound 
on the mass of $B^{(1)}$) and $\Lambda/\mu \geq 2$, 
$\sin(\theta_{KK}) < 0.15$, 
and to the accuracy required here we can take 
$\gamma^{(1)} \simeq B^{(1)}$, and $Z^{(1)} \simeq W_3^{(1)}$.

\subsection{Scalars in the Electroweak Sector} 
\label{ewsector}

We have just shown that in the absence of other scalar interactions 
$A_5^{(n)}$ 
is the Goldstone boson eaten by $A_{\mu}^{(n)}$ in the effective 
4D theory. 
In the weak sector, however, the interactions between $A_5$ and the KK 
scalar fields mix up the mass eigenstates.  Note that this was
also discussed in Ref.~\cite{DeCurtis:2002nd}.

To see this, consider the 2-point vertices of the SU(2) gauge bosons. The 
5D gauge field strength contributes a term of the form (\ref{GKin}).
Now we add to this the interactions with the standard model 
scalar fields:
\begin{eqnarray}
L_{scalar}&=&\frac{1}{2}\left|\partial_m \Phi - i g A_m^a T^a \Phi\right|^2  
\nonumber \\
& = &\frac{1}{2} \Big\{ \partial_m \Phi \partial^m \Phi+i g A_{m}^a 
\left[(T^a \Phi)^\dag \partial_m \Phi - \partial_m \Phi^\dag T^{a}\Phi 
\right] \nonumber \\
& &{} \;\; +g^2 A_m^a (T^a \Phi)^\dag A^m _b (T^b \Phi) \Big\}
\end{eqnarray}
where $A_m$, $\Phi$ are the five-dimensional fields.  We now will determine 
the physical and Goldstone scalars by examining all of the 2-point 
vertices in this Lagrangian. 

After integrating out 
the $5^{th}$ dimension, and adding in the gauge boson 
kinetic term, the relevant piece of the Lagrangian is:
\begin{eqnarray} \label{pt2}
m_{KK} A_{\mu a}^{(n)} \partial^{\mu} a_a^{(n)}  
 -\frac{1}{2} |\Phi|^2  {m_{{{\it KK}}}}^{2} +
\frac{g}{2}A^{(n)}_{\mu a} \left[ T^a \overline{v} \partial^\mu \Phi^{(n)*} 
-\partial^\mu \Phi^{(n)} (T^a \overline{v})^* \right] \nonumber \\
+ \frac{g}{2} a_a^{(n)} \left[ T^a 
\overline{v} m_{KK} \Phi^{(n)*} -m_{KK} \Phi^{(n)} (T^a \overline{v})^* \right]
\nonumber \\
+ \frac{g^2}{2} 
A_a^{\mu (n)} A_{b \mu}^{(n)} \left[ (T^a v)^\dag T^b v +m_{KK}^2 \right]
- a_a{(n)} a_b^{(n)} (T^a v)^\dag T^b v 
\end{eqnarray}
where $\overline{v} = (0,v)$ is the standard model Higgs VEV.  The terms 
involving $T^a  \overline{v} \partial_M\Phi$ give 2-point couplings between 
the KK weak gauge boson and the 
KK excitation of the corresponding Goldstone boson, which we write 
as $\Phi_a^{(n)}$.  In the case of $A_5^a (\equiv a^a)$, these 2-point 
couplings involve $p_5 \equiv m_{KK}$ rather than spatial derivatives, 
and thus become mass mixing terms between $\Phi^a$ and $a^a$.  
In addition, both $\Phi$ and $A_\mu^a$ acquire KK mass from their 
interactions with $a^a$, while $a^a$ gets a mass from its interactions 
with the Higgs VEV.  Thus $a^a$ is no longer purely a Goldstone boson, 
as it has an electroweak scale mass.

A convenient gauge-fixing functional is:
\begin{equation}
G= \frac{1}{\sqrt{\xi}} \left[ \partial^\mu A_{\mu a}^{(n)} 
- \xi( g F^a \Phi_a^{(n)} + m_{KK} a^{(n)}_a )\right]
\end{equation}
After fixing the gauge via $L \rightarrow L -\frac{1}{2} G^2$, 
all 2-point interactions between the scalars and gauge bosons 
of the effective 4D theory are canceled for any gauge fixing 
parameter $\xi$.  This means that one is free 
to choose a gauge ($\xi\rightarrow \infty $) in which the Goldstone boson 
propagator vanishes.  For general values of $\xi$, there are 
mass mixing terms between $a_a^{(n)}$ and $\Phi_a^{(n)}$. 
This leads to mass matrices for the scalars of the form
\begin{equation}
\left( \begin{array}{cc} 
 m_{\it KK}^2 \xi + m_{z}^2           & -m_Z m_{\it KK} \left( \xi-1 \right) \\
 -m_Z m_{\it KK} \left( \xi-1 \right) & m_Z^2 \xi + m_{\it KK}^2
\end{array} \right) 
\end{equation}
between $a^a$ and the KK excitation of the corresponding SM Goldstone 
boson.  Here $m_W$ and $m_Z$ are the masses of the SM 
$W$ and $Z$ gauge bosons, respectively.  

To obtain the physical and Goldstone particles, we diagonalize the mass 
matrix.\footnote{To avoid potential confusion with signs, note that the signs 
of the off-diagonal terms in the mass matrix depend on how we define 
$\Phi$ and $\Phi^*$. 
For the above, we have used the parameterization $ \Phi =\left (i 
(w_1+iw_2), v+ \phi - iz \right)$; interchanging $\Phi$ and $\Phi^*$ in 
this convention will change the sign of the off-diagonal terms in the mass 
matrix, which introduces a relative $-$ sign between the $a$ and $\Phi$ 
components of the physical scalar.  In order to ensure that the  
cancellations necessary to maintain the unitarity of the theory 
occur, care must be taken to use the correct sign 
conventions when computing vertex couplings.}
We find the mass eigenstates
\begin{eqnarray*}
\rho^{(1)} &=& \frac{1}{\sqrt{m_{z}^2+m_{KK}^2}} (-m_{KK}a + m_{z} \Phi)
  \qquad \mbox{with mass} \quad \xi \sqrt{m_{KK}^2+m_{z}^2} \\
\rho^{(2)} &=& \frac{1}{\sqrt{m_{z}^2+m_{KK}^2}} (m_{z}a+  m_{KK}\Phi)
  \;\;\> \qquad \mbox{with mass} \quad \sqrt{m_{KK}^2+m_z^2} \\
\end{eqnarray*}
Clearly $\rho^{(1)}$ is the Goldstone boson, which we can be
eliminated from the theory by taking the 
limit $\xi \rightarrow \infty$, leaving $\rho^{(2)}$ the physical scalar 
particle.  The fact that the physical scalar is a mixture of $A_{5a}$ 
and $\Phi_a$ changes its couplings to other particles by terms proportional 
to the masses of the electroweak gauge bosons.  Note that this mixing
in the scalars is needed to ensure unitarity of gauge boson scattering 
is preserved, which we explicitly verified.

Since the $A_5$ component of the physical scalar is suppressed by 
$\frac{m_Z}{m_{KK}}$ relative to the $\Phi^{(1)}$ component, in practice 
this mixing can be ignored in all calculations not involving massive 
external SM gauge bosons.

This completes our discussion of the particle spectrum of the UED model.  
The KK zero modes are the particles of the Standard Model.  
At $n_{KK}=1$ and higher, the effective theory contains massive 
vector bosons that have eaten the corresponding Goldstone $A_5^{(n)}$ 
(or a combination of $A_5^{(n)}$ and $\Phi^{(n)}$ for the $W$ and $Z$).  
It also contains both helicities of SU(2) doublet and singlet fermions, 
with the SM helicity even under the $Z_2$ action, and the non-SM helicity 
odd.  Finally, the model contains physical scalars, which are the KK Higgs, 
$h^{(n)}$, and $\phi^{(n)}_W, \phi^{(n)}_Z$, that are mixtures of 
$A_5^{W,Z}$ with the KK excitations of the SM Goldstone bosons $\phi_{W,Z}$.

\section{Feynman Rules}
\label{FeynmanRules}

In this section the Feynman rules relevant for the calculations used 
in this paper are written.  Only the relevant interactions, namely
between SM particles and level one KK modes, are shown.  
In the diagrams we use double and single lines to denote KK particles 
and SM particles, respectively.

\subsection{Fermion/Gauge Boson Interactions}

The fermion interactions of the KK modes differ from those of the 
standard model due to the vector-like nature of these higher modes, 
which introduces helicity operators at certain vertices. 
We find it convenient to work in unitary gauge, so that the Goldstone 
bosons do not appear as external particles 
but instead as the longitudinal polarizations of the massive 
KK gauge bosons.  The relevant fermion interactions, after integrating 
over the fifth dimension, are
\begin{eqnarray}
\frac{i g}{\sqrt{\pi R}} \left [\overline{f^{(0)}} \gamma^{\mu} A_\mu^{(1)} P_L
F_L^{(1)} + \overline{P_LF_L^{(1)}} 
\gamma^\mu A_\mu^{(1)} f^{(0)} \right ]\\
\frac{ig}{\sqrt{\pi R}} \left [\overline{P_L F^{(1)}_L} \gamma^{\mu} 
A_\mu^{(0)} P_LF_L^{(1)} 
 +\overline{P_RF_R^{(1)}}  \gamma^{\mu} A_\mu^{(0)}  P_RF_R^{(1)} \right ] 
\end{eqnarray} 
leading to the following vertices:
\begin{center}
\begin{picture}(300,60)(0,0)
\Text(10,60)[]{\large $f^{(0)} F^{(1)} A_\mu^{(1)}$}
\Vertex(20,25){1.5}
\ArrowLine(0,0)(20,25)
\Line(02,0)(22,25)
\ArrowLine(20,25)(00,50)
\Photon(20,25)(50,25){3}{6}
\Photon(18,25)(48,25){3}{6}
\Text(230,35)[]{\large  $i g \gamma^\mu P_L$ (ingoing $f$ or outgoing $F$)} 
\Text(230,15)[]{\large  $i g \gamma^\mu P_R$ (ingoing $F$ or outgoing $f$)} 
\end{picture}
\end{center}
\begin{center}
\begin{picture}(300,60)(0,0)
\Text(10,60)[]{\large $F^{(1)} F^{(1)} A_\mu^{(0)}$}
\Vertex(20,25){1.5}
\ArrowLine(0,0)(20,25)
\Line(02,0)(22,25)
\ArrowLine(20,25)(00,50)
\Line(22,25)(02,50)
\Photon(20,25)(50,25){3}{6}
\Text(230,35)[]{\large  $i g \gamma^\mu$} 
\end{picture}
\end{center}
where $\overline{F_L} P_L F_L +\overline{F_R} P_R F_R = \overline{F} F$;
i.e., this vertex is non-chiral as expected.

\subsubsection{Fermion/Scalar Interactions}

The interactions of fermions with scalars proportional to Yukawa 
couplings are presented for completeness, even though we do not make
use of them since we have ignored terms of order $v^2/m_{KK}^2$.
The Yukawa terms result in
\begin{eqnarray}
\overline{f}^{(0)}
\left [\frac{ \lambda_{u}}{\sqrt{\pi R}} P_R G_{uR}^{(1)} 
i\sigma_2 \Phi^{*(1)} + \frac{\lambda_{D}}{\sqrt{\pi R}} P_R G_{DR}^{(1)} 
\Phi^{(1)} \right ] \\ 
\overline{P_L F_L^{(1)}} \left [\frac{ \lambda_{u}}{\sqrt{\pi R}} P_R G_{uR}^{(1)} 
i\sigma_2 \phi^{*} + \frac{\lambda_{D}}{\sqrt{\pi R}} P_R G_{DR}^{(1)} 
\phi \right ] \\ 
\overline{P_R F_R^{(1)}} \left [\frac{ \lambda_{u}}{\sqrt{\pi R}} P_L G_{uL}^{(1)} 
i\sigma_2 \phi^{*} + \frac{\lambda_{D}}{\sqrt{\pi R}} P_L G_{DL}^{(1)} 
\phi \right ] \\ 
\overline{P_L F_L^{(1)}} \left [\frac{ \lambda_{u}}{\sqrt{\pi R}} g_{u} 
i\sigma_2 \Phi^{*(1)} + \frac{\lambda_{D}}{\sqrt{\pi R}} g_{D} 
\Phi^{(1)} \right ]  
\end{eqnarray}
where $F$ ($f$) and $G$ ($g$) are the KK (SM) SU(2) singlet and
doublet fermions, respectively.
In addition, the interactions of $A_5$ with fermions can be deduced from 
the gauge boson/fermion interaction Lagrangian above.  The only 
difference is that $A_5$ is odd under the orbifold $Z_2$, and so
it couples to the ``wrong'' (opposite) handedness of the KK fermions.  
After integrating over the fifth dimension, the interaction terms become:
\begin{eqnarray}
\frac{i g}{\sqrt{\pi R}} \left [\overline{f^{(0)}} \gamma^{5} A_5^{(1)} P_L 
F_R^{(1)} + \overline{P_L F_R^{(1)}} 
\gamma^5 A_5^{(1)} f^{(0)} \right ]
\end{eqnarray} 
and similarly for $G$, replacing left with right KK fermions.
This leads to the Feynman rules:

\begin{center}
\begin{picture}(300,60)(0,0)
\Text(10,60)[]{\large $f^{(0)} F^{(1)} A_5^{(1)}$}
\Vertex(20,25){1.5}
\ArrowLine(0,0)(20,25)
\Line(02,0)(22,25)
\ArrowLine(20,25)(00,50)
\DashLine(20,25)(50,25){4}
\DashLine(18,25)(48,25){4}
\Text(230,35)[]{\large  $\frac{g}{\sqrt{\pi R}} P_R$ (for $F$) } 
\Text(230,15)[]{\large  $\frac{-g}{\sqrt{\pi R}} P_L$ (for $G$) } 
\end{picture}
\end{center}
\begin{center}
\begin{picture}(300,60)(0,0)
\Text(10,60)[]{\large $f^{(0)} G^{(1)} \Phi^{(1)}$}
\Vertex(20,25){1.5}
\ArrowLine(0,0)(20,25)
\Line(02,0)(22,25)
\ArrowLine(20,25)(00,50)
\DashLine(20,25)(50,25){4}
\DashLine(18,25)(48,25){4}
\Text(230,35)[]{\large $\frac{\lambda_u}{\sqrt{\pi R}} i\sigma_2 P_R$ 
(for $G_u, \Phi^*$)} 
\Text(230,15)[]{\large $\frac{\lambda_D}{\sqrt{\pi R}} P_R$ (for $G_D, \Phi$)}
\end{picture}
\end{center}
\begin{center}
\begin{picture}(300,60)(0,0)
\Text(10,60)[]{\large $F^{(1)} g^{(0)} \Phi^{(1)}$}
\Vertex(20,25){1.5}
\ArrowLine(0,0)(20,25)
\Line(02,0)(22,25)
\ArrowLine(20,25)(00,50)
\DashLine(20,25)(50,25){3}
\DashLine(18,25)(48,25){3}
\Text(230,35)[]{\large $\frac{\lambda_u}{\sqrt{\pi R}} i\sigma_2 P_L$ 
(for $g_u, \Phi^*$) } 
\Text(230,15)[]{\large $\frac{\lambda_D}{\sqrt{\pi R}} P_L$ (for $g_D, \Phi$)}
\end{picture}
\end{center}
\begin{center}
\begin{picture}(300,60)(0,0)
\Text(10,60)[]{\large $F^{(1)} F^{(1)} \phi^{(0)}$}
\Vertex(20,25){1.5}
\ArrowLine(0,0)(20,25)
\Line(02,0)(22,25)
\ArrowLine(20,25)(00,50)
\Line(22,25)(02,50)
\DashLine(20,25)(50,25){3}
\Text(230,35)[]{\large $\frac{\lambda_u}{\sqrt{\pi R}} i \sigma_2$ (for $G_u, 
\phi^*$)} 
\Text(230,15)[]{\large $\frac{\lambda_D}{\sqrt{\pi R}}$ (for $G_D, \phi$)} 
\end{picture}
\end{center}
where $\overline{F_L} P_R G_R + \overline{F_R} P_L G_L = \overline{F} G$.
Unlike the vertices between SM scalars and fermions, which are suppressed 
by the small Yukawa couplings, the intrinsic strength of the vertex 
between $A_5$ and fermions is not small.  However, since the physical 
scalar particle is $m_Z A_5 + m_{KK} \Phi$, in the limit 
that that both fermion masses and $m_Z$ are small compared with $m_{KK}$, 
we can ignore all scalar-fermion vertices.

\subsubsection{Gauge Boson/Scalar Interactions}

The electroweak gauge boson/scalar interactions are of the form
\begin{eqnarray}\label{ewvertices}
\frac{ig}{\sqrt{\pi R}} \left[ 
A^{(0)}_\mu \left(\Phi \partial^\mu \Phi^* - \Phi^* \partial^\mu \Phi \right) +
A^{(1)}_\mu \left(\phi \partial^\mu \Phi^* - \Phi^* \partial^\mu \phi 
    + \phi^* \partial^\mu \Phi - \Phi \partial^\mu \phi^* \right) \right] + 
\nonumber \\
\frac{g^2}{\pi R} \left[ 
A^{(1)}_\mu A^{(0)\mu} \left(\Phi \phi^* +\Phi^* \phi + v \Phi + v \Phi^* 
    \right) +
A^{(1)}_\mu A^{(1)\mu} \left( |\phi|^2 + v \phi + v \phi^* \right) +
A^{(0)}_\mu  A^{(0)\mu} |\Phi|^2 \right]
\end{eqnarray}
where the gauge group indices and generators have been suppressed.  
The form of these interactions are nevertheless identical to those
in the Standard Model.

Since $A_5$ and the Higgs mix with each other, as we showed above, 
it is convenient to compute the interactions of $A_5$ with electroweak 
gauge bosons and scalars.  
In the electroweak sector, where $f^{abc} = i \varepsilon^{abc}$, 
these interactions arise from
\begin{eqnarray}
ig(\partial^{\mu} A^a_5 -\partial_5 A^{a\mu} )
(A^b_{\mu}A^c_5 -A^b_5 A^c_\mu) 
-\frac{g^2}{2}(A^b_{\mu}A^c_5 -A^b_5 A^c_\mu)^2 \; .
\end{eqnarray} 
After integrating over the fifth dimension, the only 
terms which survive are (using the notation $a_k$ to denote ${A_5}_k$):
\begin{equation}
\left( -\partial^\mu A^{a(n)}_5 
m^{(n)}_{KK} A^{a \mu(n)}\right) \left(A^{b (0)}_{\mu} A^{c(n)}_5 - 
A^{b(n)}_5 
A^{c(0)}_{\mu} \right) -\frac{g^2}{2}(A^{b(0)}_{\mu}A^{c(n)}_5 -A^{b(n)}_5 
A^{c(0)}_\mu)^2 
\end{equation}
Summing over $a,b,c \in SU(2)$ and switching into the usual $W^+, W^-, A, 
Z$ basis gives the following 3-point interaction terms:
\begin{eqnarray}
&i \left( -{\it W^{+(n)}}_{{{\it \nu}}}
{\it 
W^{-}}_{{\nu}}+{\it W^{-(n)}
}_{{{\it \nu}}}
{\it W^+}_{{\nu}} \right) m_{{{\it KK}}}({\it s}_{{w}}
a_{{\gamma}} + c_{w} 
a_{{Z}} )\nonumber \\
&+i \left( {\it W^{+ (n)}}_{{{\nu}}}(Z_{{\nu}}c_{{w}}+
A_{{\nu}}{\it s}_{{w}})-(A^{(n)}_{{{\nu}}}{\it s}_{{w}}
+Z^{(n)}_{{{\nu}}}c_{{w}}){\it W^+}_{{\nu}} \right) m_{{{\it 
KK}}}a_{{-}} \nonumber \\
&+i \left( -{\it W^{-(n)}}_{{{\nu}}}(Z_{{\nu}}c_{{w}}+A_{{\nu}}{\it s}_{{w}}) 
+(Z^{(n)}_{{{\nu}}}c_{{w}}
+A^{(n)}_{{{\nu}}}{\it s}_{{w}}){\it 
W^-}_{{\nu}}\right) 
m_{{{\it 
KK}}}a_{{+}} 
\end{eqnarray}
for the two gauge boson/$A_5$ interactions, and
\begin{eqnarray}
&-i \left( a_{{-}}^{(n)} {\partial} a_{{+}}^{(n)}
 -a^{(n)}_{{+}}{\partial} a_{{-}}^{(n)}  \right) {\it s}_w A^{(0)}_{{
\nu}}-i \left( -a_{{+}}^{(n)}{\partial} a_{{-}} +a_{{-}}^{(n)}
{\partial} a_{{+}}^{(n)}  \right) c_w Z_{{\nu}}^{(0)}
 \nonumber \\
&-i \left( \left \{a_{{z}}^{(n)}c_{{w}} +a_{{\gamma
}}^{(n)}{\it s}_{{w}}\right \} {\partial} a_{{-}}^{(n)}
  -a_{{-}}^{(n)} \left \{{\partial}  a_{{\gamma}}^{(n)}
 {\it s}_{{w}}{\partial} a_{{z}}^{(n)}
 c_{{w}}\right\} \right) {\it W^{+(0)}}_{{\nu}}- \nonumber \\
&i \left( 
 -\left \{a_{{z}}^{(n)}c_{{w}} +a_{{\gamma}}^{(n)}
{\it s}_{{w}} \right \}{\partial}
 a_{{+}}^{(n)} +a_{{+}}^{(n)}
\left \{{\partial} a_{{\gamma}}^{(n)}
 {\it s}_{{w}}+{\partial} a_{{z}}^{(n)} c_{{w}}\right \}
\right) {\it W^{-(0)}}_{{\nu}} 
\end{eqnarray}
for the triple-scalar/gauge boson interactions.
The 4-point interactions are obtained from the corresponding standard 
model vertices by requiring both KK particles be $A_5$.

$A_5$ also couples to $\Phi$ through $|D_{\mu} \Phi |^2$.  This results in 
three types of vertices: $V_1 = g m_{KK} A_5 \Phi \Phi^*$, $V_2= g m_Z A_5 
A_5 \Phi$, 
and  $V_3= g^2 A_5 A_5 \Phi \Phi^*$.  
$V_2$ and $V_3$ are identical to 
those of the KK gauge bosons, once the factors of 
$g_{\mu \nu}$ are replaced with $g_{5 5}$.  $V_1$ is obtained from the 
corresponding KK gauge boson vertex by replacing $p_\mu \rightarrow i 
m_{KK}$. As some care must be taken with signs, 
the $V_1$ terms are listed here:
\begin{eqnarray}
-i{\it s}_{{w}} \left( -{w_{{+}}}^{(n)}w_{{-}}+{w_{{-}}}^{(n)}w_{{+}}
 \right) a_{{\gamma}}^{(n)} 
\nonumber \\
+\frac{1}{2\cos \left( \theta_w \right) 
 }\,a_{{z}}^{(n)} \left( \,i  \cos \left( 2 
\theta_w 
 \right)({w_{{+}}}^{(n)}w_{{-}}-
{w_{{-}}}^{(n)}w_{{+}} )
+{h}^{(n)}z-{z}^{(n)}h \right) 
 \nonumber \\
+\frac{1}{2}\, \left( 
-iz{w_{{+}}}^{(n)}-h{w_{{+}}}^{(n)}+{h}^{(n)}w_{{+}}+i{z}^{(n)}w_
{{+}} \right) a_{{-}} \nonumber \\
+\frac{1}{2}\, \left( 
iz{w_{{-}}}^{(n)}-h{w_{{-}}}^{(n)}+{h}^{(n)}w_{{-}}-i{z}^{(n)}w_{
{-}} \right) a_{{+}}
\end{eqnarray}

The vertices involving physical scalars follow from combining vertices 
involving $A_5$ and $\Phi$ in the correct proportion.  A table of these 
physical vertices is given below.  In principle, care must be taken 
in vertices involving the Weinberg angle, as the mass mixing is not 
the same for the $KK$ particles as it is for the SM particles.
In the case of the scalar $A_5$, however, the physical component of 
$A_5$ is that which mixes with the SM Goldstone boson of the $Z$ particle, 
and hence its mixing angle should be the same as that of the SM $Z$ boson.  
For the KK gauge bosons, the mixing angle is different.  However, as we 
already discussed above, it is a good approximation to neglect
this mixing and thus take
$\gamma^{(1)} \simeq B^{(1)}$ and $Z^{(1)} \simeq W^{3(1)}$.

The following notation is used in the table.  We define 
$\kappa_w = \frac{m_{KK}}{\sqrt{m_{KK}^2+m_W^2}}$ and $\kappa_z = 
\frac{m_{KK}}{\sqrt{m_{KK}^2+m_Z^2}}$.  
$\Phi_z, \Phi_{+}$, and 
$\Phi_{-}$  are the 
physical level one KK scalars that are themselves mixtures of the 
KK excitations of the SM Goldstone bosons and $A_5$ from the weak 
gauge bosons, as described in Sec.~\ref{ewsector}.  
(Here we use level one KK particles, but the vertices 
are the same for level $n$ KK states also.).  
$\Phi_h$ is the level one KK Higgs particle, 
while $h$ is the SM Higgs particle.  The table includes only vertices 
which differ from those of the SM.

\begin{center}
\begin{tabular} {|c|c|}
\multicolumn{2}{c}{2 vector/scalar vertices} \\
\hline
Vertex &  Coupling \\
$W^{+(1)}_\mu W^{- (0)}_\nu \Phi_z$ &$ -i c_w (-1)^\beta 
m_{z}\kappa_z g_{\mu 
\nu} $\\
$W^{-(1)}_\mu W^{+ (0)}_\nu \Phi_z$ &$ i c_w(-1)^\beta  
m_{z} \kappa_z g_{\mu \nu} 
$\\
$W^{+(1)}_\mu Z^{ (0)}_\nu \Phi_-$ &$ i m_Z (s_w^2+(-1)^\beta c_w^2)
\kappa_w g_{\mu \nu} $\\
$W^{-(1)}_\mu Z^{ (0)}_\nu \Phi_+$ &$ -i m_Z
(s_w^2+(-1)^\beta c_w^2)
 \kappa_w  g_{\mu \nu} $\\
$W^{-(0)}_\mu Z^{ (1)}_\nu \Phi_+$ &$ -i m_Z (s_w^2-(-1)^\beta c_w^2)
 \kappa_w  g_{\mu \nu} $\\
$W^{+(0)}_\mu Z^{ (1)}_\nu \Phi_-$ &$ i m_Z (s_w^2-(-1)^\beta c_w^2)
 \kappa_w  
g_{\mu \nu} $\\
$W^{+(1)}_\mu A^{ (0)}_\nu \Phi_-$ & 
 $\frac{1}{2} i m_Z \sin(2 \theta _w) 
((-1)^\beta -1) 
\kappa_w  g_{\mu \nu} $\\
$W^{-(1)}_\mu A^{ (0)}_\nu \Phi_+$ & $\frac{-1}{2} i m_Z \sin(2 \theta _w) 
((-1)^\beta -1) 
\kappa_w  g_{\mu \nu} $\\
$W^{-(0)}_\mu A^{ (1)}_\nu \Phi_+$ &$ 
\frac{1}{2} i m_Z \sin(2 \theta _w)((-1)^\beta +1) 
 \kappa_w  g_{\mu \nu} $\\
$W^{+(0)}_\mu A^{ (1)}_\nu \Phi_-$ &$ \frac{-1}{2}i m_Z \sin(2 \theta _w) 
((-1)^\beta -1) \kappa_w g_{\mu \nu} $\\
$A_\mu^{(i)} A_\nu ^{(j)} \Phi_h $ & as in standard model \\
\hline
\end{tabular}
\begin{tabular}{|c|c|}
\multicolumn{2}{c}{2 scalar/vector vertices} \\
\hline
Vertex & Coupling \\
$A_\mu^{(0)} \Phi_+ \Phi_- $& $g s_w(p_{+}-p_{-})$\\
$Z_\mu^{(0)} \Phi_+ \Phi_- $& $ g(c_w- \frac{cos(2\theta_W)}{2 
c_w}\kappa_w^2) 
(p_{+}-p_{-})$\\
$W_\mu^{+(0)} \Phi_z \Phi_- $& $ g\left ( \frac{m_W^2}{m_{KK}^2} 
+ \frac{1}{2}\right) \kappa_z \kappa_w (p_{-}-p_{z})$\\
$W_\mu^{-(0)} \Phi_z \Phi_+ $& $ -g\left ( \frac{m_W^2}{m_{KK}^2} 
+ \frac{1}{2}\right) \kappa_z \kappa_w (p_{+}-p_{z})$\\
$A_\mu^{(0)j} \Phi_j \Phi_h$ & $\kappa_j \times$ SM vertex \\
\hline
\end{tabular}
\begin{tabular} {|c|c|}
\multicolumn{2}{c}{2 vector/2 scalar vertices} \\
\hline
$ W^{+(0)}_\mu W^{-(0)}_\nu  \Phi_+ \Phi_-$ & $ \frac{g^2}{2}(1+ 
\frac{2 m_W^2}{m_{KK}^2}) \kappa_w^2  g_{\mu \nu}$ \\
$ W^{+(0)}_\mu W^{-(0)}_\nu \Phi_z \Phi_z$ & $ \frac{g^2}{2}\kappa_z^2(1+4 
\frac{c_w^2 m_Z^2}{m_{KK}^2} ) g_{\mu \nu} $ \\ 
$ A^{(0)}_\mu A^{-(0)}_\nu \Phi_+ \Phi_-$ & $ 2 e^2 g_{\mu \nu }$ \\
$ Z^{(0)}_\mu Z^{(0)}_\nu  \Phi_+ \Phi_-$ & $ g^2 \kappa_w^2(
\frac{\cos(2 \theta_w)^2}{2 c_w^2}+ \frac{2 c_w^4 m_Z^2}{m_{KK}^2} ) g_{\mu 
\nu}$ \\
$ Z^{(0)}_\mu A^{(0)}_\nu  \Phi_+ \Phi_-$ & $ ge \kappa_w^2( \frac{\cos(2 
\theta_w)}{c_w} +\frac{2 c_w^3 m_Z^2}{m_{KK}^2}) g_{\mu \nu}$ \\
$ Z^{(0)}_\mu W^{\pm (0)}_\nu  \Phi_z \Phi_\pm$ & $ \frac{g^2}{2} 
\kappa_w \kappa_z( \frac{s_w^2}{c_w} -\frac{c_w^3 m_Z^2}{m_{KK}^2} )
g_{\mu \nu}$ \\
$ A^{(0)}_\mu W^{\pm (0)}_\nu  \Phi_z \Phi_\pm$ & $ \frac{-ge}{2} 
\kappa_w \kappa_z( 1 +\frac{c_w^2 m_Z^2}{m_{KK}^2} )g_{\mu \nu}$ \\
\hline
\end{tabular}
\begin{tabular} {|c|c|}
\multicolumn{2}{c}{3 scalar vertices} \\
\hline
$\Phi_z \Phi_z h$& $ 
-(m_Z^2(1+\frac{m_Z^2}{m_{KK}^2})+\frac{1}{2}M_h^2) 
\frac{\kappa_z^2}{c_w m_Z} $ \\ 
$\Phi_w \Phi_w h$& $ 
-(m_W^2(1+\frac{m_W^2}{m_{KK}^2})+\frac{1}{2}M_h^2) 
\frac{\kappa_w^2}{c_w m_Z} $ \\ 
\hline
\multicolumn{2}{c}{4 scalar vertices} \\
\hline
$\Phi_+ \Phi_- h h$ &$ \frac{-g^2 \kappa_w^2}{4 } 
(\frac{M_h^2}{m_W^2} 
+ 2 \frac{m_W^2}{m_{KK}^2} )$ \\
$\Phi_z \Phi_z h h$ & $\frac{-g^2 \kappa_z^2}{4 c_w^2} 
(\frac{M_h^2}{m_Z^2} 
+ 2 \frac{m_Z^2}{m_{KK}^2}) $ \\
\hline
\end{tabular}
\end{center}

Here we have included only vertices in which the SM scalars are the 
physical Higgs particle.  
In practice, to lowest order in $v/m_{KK}$, the only vertices 
that are affected are those whose analogue in the SM contains a factor of 
$m_Z$.  In this case, the contributions of certain $A_5$ terms can be of 
the same order in $m_Z$ as the contributions of $\Phi^{(1)}$.

Note that vertices of the form $A^{\mu a} A^b_{\mu} a^c$ have 
a sign which depends on whether the indices $a,b,c$ occur in clockwise 
or counter-clockwise order, caused by $\varepsilon^{a b c}$ from the 
commutators of $SU(2)$ generators.  Diagrams of the form $A^{a \mu} 
A^a_{\mu} \Phi$, have no such sign.  Hence care must be taken in 
combining the contributions of these two types of diagrams to obtain the 
physical scalar vertex.  In order to make this clear, we have explicitly 
kept a factor of $(-1)^\beta$ in the relevant vertices.  Here $\beta$ is 
even for vertices in which the electric charge entering the vertex 
increases in the counter-clockwise direction, and is odd otherwise.

The gauge boson self-interaction terms are exactly as in the standard 
model, and need not be reviewed here.  Note that $A^{(1)}_{\mu}$, 
being even under orbifold parity, has no vertices with the Goldstone 
boson $A_5$ apart from those in the kinetic terms described above.

\section{Generic Diagrams and their annihilation cross sections}  
\label{Diagrams}

In this section, we list all Feynman diagrams involved in the annihilation 
processes we consider.  (As explained above, processes involving the 
Yukawa coupling of Higgs to fermions, the self-coupling of Higgs to 
itself, are not calculated as they are expected to have a small effect on 
these results.)  The possible annihilation processes are classified 
according to the nature of the initial and final state particles.  Some 
diagrams apply to multiple 
processes, and not all diagrams in a given section necessarily apply to 
all processes of that type.  A 
comprehensive list of processes, the relevant diagrams, and the 
corresponding cross sections can be found in Appendix~\ref{CrossSect}.  

\subsection{Fermion Annihilation}

\subsubsection{$f^{(1)} f^{(1)} \rightarrow f^{(0)} f^{(0)}$ and $f^{(1)} 
\overline{f}^{(1)} \rightarrow f^{(0)} \overline{f}^{(0)}$}

\begin{picture}(400,50)(0,0)
\Text(0,35)[]{\large a}
\Vertex(80,25){1.5}
\Vertex(20,25){1.5}
\ArrowLine(0,0)(20,25)
\Line(02,0)(22,25)
\ArrowLine(100,0)(80,25)
\Line(98,0)(78,25)
\ArrowLine(20,25)(00,50)
\ArrowLine(80,25)(100,50)
\Photon(20,25)(80,25){3}{6}
\Photon(20,24)(80,24){3}{6}

\Text(150,35)[]{\large b}
\Vertex(230,25){1.5}
\Vertex(170,25){1.5}
\ArrowLine(150,0)(170,25)
\Line(152,0)(172,25)
\Line(248,0)(228,25)
\ArrowLine(250,0)(230,25)
\ArrowLine(230,25)(150,55)
\ArrowLine(170,25)(250,55)
\Photon(170,25)(230,25){3}{6}
\Photon(170,24)(230,24){3}{6}

\Text(300,35)[]{\large c}
\Vertex(380,25){1.5}
\Vertex(320,25){1.5}
\ArrowLine(300,0)(320,25)
\Line(302,0)(322,25)
\ArrowLine(380,25)(400,0)
\Line(398,0)(378,25)
\ArrowLine(320,25)(300,50)
\ArrowLine(400,50)(380,25)
\Photon(320,25)(380,25){3}{6}
\Photon(320,24)(380,24){3}{6}

\Text(430,55)[]{\large d}
\Vertex(450,25){1.5}
\Vertex(450,50){1.5}
\ArrowLine(420,0)(450,25)
\Line(422,0)(452,25)
\Line(478,0)(448,25)
\ArrowLine(450,25)(480,0)
\ArrowLine(450,50)(420,75)
\ArrowLine(480,75)(450,50)
\Photon(450,25)(450,50){3}{3}
\end{picture}

\subsubsection{$f^{(1)} \overline{f}^{(1)} \rightarrow 
               A_{\mu}^{(0)} A_{\nu}^{(0)}$}

Here the set of diagrams depends on the particular final state;
i.e., the last diagram is absent when the final state gauge bosons are
hypercharge.

\begin{picture}(450,60)(0,0)
\Text(0,35)[]{\large a}
\Vertex(80,25){1.5}
\Vertex(20,25){1.5}
\ArrowLine(0,0)(20,25)
\Line(02,0)(22,25)
\ArrowLine(80,25)(100,0)
\Line(98,0)(78,25)
\Photon(20,25)(00,50){3}{4}
\Photon(100,50)(80,25){3}{4}
\ArrowLine(20,26)(80,26)
\Line(80,24)(20,24)

\Text(200,35)[]{\large b}
\Vertex(280,25){1.5}
\Vertex(220,25){1.5}
\ArrowLine(200,0)(220,25)
\Line(202,0)(222,25)
\Line(298,0)(278,25)
\ArrowLine(280,25)(300,0)
\Photon(280,25)(200,55){3}{8}
\Photon(220,25)(300,55){3}{8}
\ArrowLine(220,26)(280,26)
\Line(280,24)(220,24)

\Text(370,55)[]{\large c}
\Vertex(400,25){1.5}
\Vertex(400,50){1.5}
\ArrowLine(370,0)(400,25)
\Line(372,0)(402,25)
\Line(428,0)(398,25)
\ArrowLine(400,25)(430,0)
\Photon(400,50)(370,75){3}{4}
\Photon(430,75)(400,50){3}{4}
\Photon(400,25)(400,50){3}{3}
\end{picture}

\subsubsection{$f^{(1)} \overline{f}^{(1)} \rightarrow \phi^{(0)} \phi^{*(0)}$}

\begin{picture}(400,80)(0,0)
\Vertex(50,25){1.5}
\Vertex(50,50){1.5}
\ArrowLine(20,0)(50,25)
\Line(22,0)(52,25)
\Line(78,0)(48,25)
\ArrowLine(50,25)(80,0)
\DashLine(50,50)(20,75){4}
\DashLine(80,75)(50,50){4}
\Photon(50,25)(50,50){3}{3}
\end{picture}

\subsection{Gauge Boson Annihilation}

\subsubsection{$A_{\mu}^{(1)} A_{\nu}^{(1)} \rightarrow 
               f^{(0)} \overline{f}^{(0)} $}

Here vertices involving both SM and KK fermions are chiral, as previously 
mentioned, and factors of the appropriate helicity projection operators 
must be included.  In the limit that electroweak breaking masses are ignored, 
all of these scattering processes result in two final state fermions 
of the same chirality.

\begin{picture}(450,70)(0,0)
\Text(0,35)[]{\large a}
\Vertex(80,25){1.5}
\Vertex(20,25){1.5}
\Photon(0,0)(20,25){3}{4}
\Photon(02,0)(22,25){3}{4}
\Photon(80,25)(100,0){3}{4}
\Photon(98,0)(78,25){3}{4}
\ArrowLine(20,25)(00,50)
\ArrowLine(100,50)(80,25)
\ArrowLine(80,26)(20,26)
\Line(80,24)(20,24)

\Text(200,35)[]{\large b}
\Vertex(280,25){1.5}
\Vertex(220,25){1.5}
\Photon(200,0)(278,25){3}{8}
\Photon(202,0)(280,25){3}{8}
\Photon(298,0)(220,25){3}{8}
\Photon(222,25)(300,0){3}{8}
\ArrowLine(220,25)(200,55)
\ArrowLine(300,55)(280,25)
\ArrowLine(280,26)(220,26)
\Line(280,24)(220,24)

\Text(370,55)[]{\large c}
\Vertex(400,25){1.5}
\Vertex(400,50){1.5}
\Photon(370,0)(400,25){3}{4}
\Photon(372,0)(402,25){3}{4}
\Photon(428,0)(398,25){3}{4}
\Photon(400,25)(430,0){3}{4}
\ArrowLine(400,50)(370,75)
\ArrowLine(430,75)(400,50)
\Photon(400,25)(400,50){3}{3}
\end{picture}

\subsubsection{$W_{\mu}^{(1)} W_{\nu}^{(1)} \rightarrow \phi^{(0)} \phi^{*(0)}$
and $Z_{\mu}^{(1)} Z_{\nu}^{(1)} \rightarrow \phi^{(0)} \phi^{*(0)}$}

This set of diagrams corresponds to the annihilation of $W$ or $Z$ 
bosons into at least one physical SM Higgs boson.  
In the limit 
that the masses of 
the SM $W$ and $Z$ are small compared to $s$,  we can consistently ignore  
diagrams with vertices of the form $WW \phi$, $ZZ \phi$, and $WZ \phi$, 
as their couplings are suppressed by $v/m_{KK}$ or $v^2/s$.  
If one of the external particles is a gauge boson, 
the Goldstone boson equivalence theorem tells us that
only the longitudinal polarization of the external gauge boson 
contributes.  We can then use the Goldstone boson approximation to 
calculate the relevant cross sections from the set of diagrams below.

\begin{picture}(450,70)(0,0)
\Text(0,35)[]{\large a}
\Vertex(80,25){1.5}
\Vertex(20,25){1.5}
\Photon(0,0)(20,25){3}{4}
\Photon(02,0)(22,25){3}{4}
\Photon(80,25)(100,0){3}{4}
\Photon(98,0)(78,25){3}{4}
\DashLine(20,25)(00,50){3}
\DashLine(100,50)(80,25){3}
\DashLine(80,26)(20,26){3}
\DashLine(80,24)(20,24){3}

\Text(150,35)[]{\large b}
\Vertex(230,25){1.5}
\Vertex(170,25){1.5}
\Photon(150,0)(228,25){3}{8}
\Photon(152,0)(230,25){3}{8}
\Photon(248,0)(170,25){3}{8}
\Photon(172,25)(250,0){3}{8}
\DashLine(170,25)(150,55){2}
\DashLine(250,55)(230,25){2}
\DashLine(230,26)(170,26){2}
\DashLine(230,24)(170,24){2}

\Text(280,35)[]{\large c}
\Vertex(320,25){1.5}
\Photon(290,0)(320,25){3}{4}
\Photon(292,0)(322,25){3}{4}
\Photon(348,0)(318,25){3}{4}
\Photon(320,25)(350,0){3}{4}
\DashLine(320,25)(290,50){2}
\DashLine(350,50)(320,25){2}

\Text(390,55)[]{\large d}
\Vertex(420,25){1.5}
\Vertex(420,50){1.5}
\Photon(390,0)(420,25){3}{4}
\Photon(392,0)(422,25){3}{4}
\Photon(448,0)(418,25){3}{4}
\Photon(420,25)(450,0){3}{4}
\DashLine(420,50)(390,75){3}
\DashLine(450,75)(420,50){3}
\Photon(420,25)(420,50){3}{3}
\end{picture}

\subsubsection{$A_{\mu}^{(1)} A_{\nu}^{(1)} \rightarrow 
               A_{\mu}^{(0)} A_{\nu}^{(0)} $}

This set of diagrams corresponds to KK gauge boson annihilation
into SM gauge bosons.  Note that we treated SM electroweak gauge bosons 
as massive, so that no external Goldstone bosons are necessary.  
However, all diagrams with physical scalar propagators must be included.  
In the $s$-channel, only the Higgs plays a role.  However, since there 
are physical KK $w$ and $z$ scalars, these propagators must be included 
in the $t$ and $u$ channels.  We remark that would-be high-energy 
divergences of the form $s/m_{KK}^2$ are absent precisely because the 
the non-Goldstone scalars are a mixture of $\phi^{(1)}$ and $A_5^{(1)}$.

\begin{picture}(450,70)(0,0)
\Text(0,35)[]{\large a}
\Vertex(80,25){1.5}
\Vertex(20,25){1.5}
\Photon(0,0)(20,25){3}{4}
\Photon(02,0)(22,25){3}{4}
\Photon(80,25)(100,0){3}{4}
\Photon(98,0)(78,25){3}{4}
\Photon(20,25)(00,50){3}{4}
\Photon(100,50)(80,25){3}{4}
\Photon(80,26)(20,26){3}{6}
\Photon(80,24)(20,24){3}{6}
\Text(230,35)[]{\large }

\Text(140,35)[]{\large b}
\Vertex(220,25){1.5}
\Vertex(160,25){1.5}
\Photon(140,0)(218,25){3}{8}
\Photon(142,0)(220,25){3}{8}
\Photon(238,0)(160,25){3}{8}
\Photon(162,25)(240,0){3}{8}
\Photon(160,25)(140,55){3}{4}
\Photon(240,55)(220,25){3}{4}
\Photon(220,26)(160,26){3}{6}
\Photon(220,24)(160,24){3}{6}

\Text(270,35)[]{\large s}
\Vertex(280,25){1.5}
\Vertex(280,50){1.5}
\Photon(250,0)(280,25){3}{4}
\Photon(252,0)(282,25){3}{4}
\Photon(308,0)(278,25){3}{4}
\Photon(280,25)(310,0){3}{4}
\Photon(280,50)(250,75){3}{4}
\Photon(310,75)(280,50){3}{4}
\Photon(280,25)(280,50){3}{3}

\Text(350,35)[]{\large c}
\Vertex(400,25){1.5}
\Photon(370,0)(400,25){3}{4}
\Photon(372,0)(402,25){3}{4}
\Photon(428,0)(398,25){3}{4}
\Photon(400,25)(430,0){3}{4}
\Photon(400,25)(370,50){3}{4}
\Photon(430,50)(400,25){3}{4}
\end{picture}

\begin{picture}(450,80)(0,0)
\Text(0,35)[]{\large e}
\Vertex(80,25){1.5}
\Vertex(20,25){1.5}
\Photon(0,0)(20,25){3}{4}
\Photon(02,0)(22,25){3}{4}
\Photon(80,25)(100,0){3}{4}
\Photon(98,0)(78,25){3}{4}
\Photon(20,25)(00,50){3}{4}
\Photon(100,50)(80,25){3}{4}
\DashLine(80,26)(20,26){3}
\DashLine(80,24)(20,24){3}

\Text(150,35)[]{\large f}
\Vertex(230,25){1.5}
\Vertex(170,25){1.5}
\Photon(150,0)(228,25){3}{8}
\Photon(152,0)(230,25){3}{8}
\Photon(248,0)(170,25){3}{8}
\Photon(172,25)(250,0){3}{8}
\Photon(170,25)(150,55){3}{4}
\Photon(250,55)(230,25){3}{4}
\DashLine(230,26)(170,26){3}
\DashLine(230,24)(170,24){3}

\Text(330,45)[]{\large d}
\Vertex(350,25){1.5}
\Vertex(350,50){1.5}
\Photon(320,0)(350,25){3}{4}
\Photon(322,0)(352,25){3}{4}
\Photon(378,0)(348,25){3}{4}
\Photon(350,25)(380,0){3}{4}
\Photon(350,50)(320,75){3}{4}
\Photon(380,75)(350,50){3}{4}
\DashLine(350,25)(350,50){3}
\end{picture}

\subsection{Higgs Annihilation}

\subsubsection{$\Phi^{(1)} \Phi^{(1)} \rightarrow \phi^{(0)} \phi^{(0)}$}

This set of diagrams correspond to scalar annihilation into at least 
one final state Higgs particle.  If final state gauge bosons are present, 
the resulting diagrams contain at least one coupling proportional to $m_Z$, 
and hence are highly suppressed unless the gauge boson is longitudinally 
polarized.  Thus all such processes can be described by purely scalar 
final states in the limit that we ignore terms proportional to $m_Z/m_{KK}$.  

Here we furthermore ignore diagrams involving 3-scalar vertices, since their 
contribution is suppressed by factors of $v/m_{KK}$ or $v^2/s$.

\begin{picture}(450,70)(0,0)
\Text(0,35)[]{\large a}
\Vertex(80,25){1.5}
\Vertex(20,25){1.5}
\DashLine(0,0)(20,25){4}  
\DashLine(02,0)(22,25){4}
\DashLine(100,0)(80,25){4}
\DashLine(98,0)(78,25){4}  
\DashLine(20,25)(00,50){4} 
\DashLine(80,25)(100,50){4}
\Photon(20,26)(80,26){3}{6}
\Photon(20,24)(80,24){3}{6}

\Text(140,35)[]{\large b}   
\Vertex(220,25){1.5}
\Vertex(160,25){1.5}
\DashLine(140,0)(160,25){4}   
\DashLine(142,0)(162,25){4}  
\DashLine(238,0)(218,25){4}  
\DashLine(240,0)(220,25){4} 
\DashLine(220,25)(140,55){4}  
\DashLine(160,25)(240,55){4}
\Photon(160,26)(220,26){3}{6}
\Photon(160,24)(220,24){3}{6}

\Text(310,55)[]{\large c}
\Vertex(340,25){1.5}
\Vertex(340,50){1.5}
\DashLine(310,0)(340,25){3}
\DashLine(312,0)(342,25){3}
\DashLine(368,0)(338,25){3}
\DashLine(340,25)(370,0){3}
\DashLine(340,50)(310,75){3}
\DashLine(370,75)(340,50){3}
\Photon(340,25)(340,50){3}{3}

\Text(390,35)[]{\large d}
\Vertex(430,25){1.5}
\DashLine(400,0)(430,25){3}
\DashLine(402,0)(432,25){3}
\DashLine(458,0)(428,25){3}
\DashLine(430,25)(460,0){3}
\DashLine(430,25)(400,50){3}
\DashLine(460,50)(430,25){3}
\end{picture}

\subsubsection{$\Phi^{(1)} \Phi^{*(1)} \rightarrow 
               A_{\mu}^{(0)} A_{\nu}^{(0)}$}

\begin{picture}(450,60)(0,0)
\Text(0,35)[]{\large a}
\Vertex(80,25){1.5}
\Vertex(20,25){1.5}
\DashLine(0,0)(20,25){4}
\DashLine(02,0)(22,25){4}
\DashLine(80,25)(100,0){4}
\DashLine(98,0)(78,25){4}
\Photon(20,25)(00,50){3}{3}
\Photon(100,50)(80,25){3}{3}
\DashLine(20,26)(80,26){4}
\DashLine(80,24)(20,24){4}

\Text(140,35)[]{\large b}
\Vertex(220,25){1.5}
\Vertex(160,25){1.5}
\DashLine(140,0)(160,25){4}
\DashLine(142,0)(162,25){4}
\DashLine(238,0)(218,25){4}
\DashLine(220,25)(240,0){4}
\Photon(220,25)(140,55){3}{8}
\Photon(160,25)(240,55){3}{8}
\DashLine(160,26)(220,26){4}
\DashLine(220,24)(160,24){4}

\Text(290,35)[]{\large c}
\Vertex(330,25){1.5}
\Photon(330,25)(300,50){3}{4}
\DashLine(302,0)(332,25){3}
\Photon(360,50)(330,25){3}{4}
\DashLine(328,25)(358,0){3}
\DashLine(300,0)(330,25){3}
\DashLine(360,0)(330,25){3}

\Text(400,55)[]{\large d }
\Vertex(430,25){1.5}
\Vertex(430,50){1.5}
\DashLine(400,0)(430,25){3}
\DashLine(402,0)(432,25){3}
\DashLine(458,0)(428,25){3}
\DashLine(430,25)(460,0){3}
\Photon(430,50)(400,75){3}{4}
\Photon(460,75)(430,50){3}{4}
\Photon(430,25)(430,50){3}{3}
\end{picture}

\begin{picture}(450,80)(0,0)
\Text(0,35)[]{\large e}
\Vertex(80,25){1.5}
\Vertex(20,25){1.5}
\DashLine(0,0)(20,25){4}
\DashLine(02,0)(22,25){4}
\DashLine(80,25)(100,0){4}
\DashLine(98,0)(78,25){4}
\Photon(20,25)(00,50){3}{4}
\Photon(100,50)(80,25){3}{4}
\Photon(20,26)(80,26){4}{6}
\Photon(80,24)(20,24){4}{6}

\Text(150,35)[]{\large f}
\Vertex(230,25){1.5}
\Vertex(170,25){1.5}
\DashLine(150,0)(170,25){4}
\DashLine(152,0)(172,25){4}
\DashLine(248,0)(228,25){4}
\DashLine(230,25)(250,0){4}
\Photon(230,25)(150,55){3}{8}
\Photon(170,25)(250,55){3}{8}
\Photon(170,26)(230,26){4}{6}
\Photon(230,24)(170,24){4}{6}

\Text(320,55)[]{\large g }
\Vertex(350,25){1.5}
\Vertex(350,50){1.5}
\DashLine(320,0)(350,25){3}
\DashLine(322,0)(352,25){3}
\DashLine(378,0)(348,25){3}
\DashLine(350,25)(380,0){3}
\Photon(350,50)(320,75){3}{4}
\Photon(380,75)(350,50){3}{4}
\DashLine(350,25)(350,50){3}
\end{picture}
 
\subsubsection{$\Phi^{(1)} \Phi^{*(1)} \rightarrow f^{(0)} \overline{f}^{(0)}$}

\begin{picture}(400,80)(0,0)
\Text(0,35)[]{\large }
\Vertex(50,25){1.5}
\Vertex(50,50){1.5}
\DashLine(20,0)(50,25){3}
\DashLine(22,0)(52,25){3}
\DashLine(78,0)(48,25){3}
\DashLine(50,25)(80,0){3}
\ArrowLine(50,50)(20,75)
\ArrowLine(80,75)(50,50)
\Photon(50,25)(50,50){3}{3}
\end{picture}

\subsection{Fermion/gauge boson coannihilation}

\subsubsection{$f^{(1)} A_{\nu}^{(1)} \rightarrow f^{(0)} A_{\nu}^{(0)}$}

\begin{picture}(450,60)(0,0)
\Text(0,35)[]{\large a}
\Vertex(80,25){1.5}
\Vertex(20,25){1.5}
\Photon(0,0)(20,25){3}{3}
\Photon(02,0)(22,25){3}{3}
\ArrowLine(100,0)(80,25)
\ArrowLine(98,0)(78,25)
\ArrowLine(20,25)(00,50)
\Photon(100,50)(80,25){3}{3}
\ArrowLine(80,26)(20,26)
\ArrowLine(80,24)(20,24)

\Text(200,35)[]{\large b}
\Vertex(280,25){1.5}
\Vertex(220,25){1.5}
\Photon(200,0)(220,25){3}{3}
\Photon(202,0)(222,25){3}{3}
\ArrowLine(300,0)(280,25)
\Line(298,0)(278,25)
\Photon(220,25)(300,50){3}{8}
\ArrowLine(280,25)(200,50)
\Photon(280,26)(220,26){3}{6}
\Photon(280,24)(220,24){3}{6}

\Text(370,45)[]{\large c}
\Vertex(400,25){1.5}
\Vertex(400,50){1.5}
\Photon(370,0)(400,25){3}{4}
\Photon(372,0)(402,25){3}{4}
\ArrowLine(428,0)(398,25)
\Line(400,25)(430,0)
\ArrowLine(400,50)(370,75)
\Photon(430,75)(400,50){3}{4}
\ArrowLine(400,25)(400,50)
\end{picture}

\subsection{Gauge boson/Higgs coannihilation}
\subsubsection{$\Phi^{(1)} A_{\nu}^{(1)} \rightarrow \Phi^{(0)} A_{\nu}^{(0)}$}

Here again we work in the limit $m_W \ll s, m_{KK}$.  In this 
limit, processes to two final state scalars can be neglected, 
as they necessarily involve at least one vertex 
suppressed by a factor of $m_Z$.  In processes to one final state scalar 
and one final state gauge boson, diagrams in which scalar propagators are 
replaced by gauge boson propagators are also electroweak suppressed 
(by $m_Z^2$, while the longitudinal gauge boson contributes 
$1/m_Z$) and thus neglected.

\begin{picture}(450,80)(0,0)
\Text(0,35)[]{\large a}
\Vertex(80,25){1.5}
\Vertex(20,25){1.5}
\Photon(0,0)(20,25){3}{3}
\Photon(02,0)(22,25){3}{3}
\DashLine(100,0)(80,25){4}
\DashLine(98,0)(78,25){4}
\DashLine(20,25)(00,50){4}
\Photon(100,50)(80,25){3}{3}
\DashLine(80,26)(20,26){4}
\DashLine(80,24)(20,24){4}

\Text(170,55)[]{\large b}
\Vertex(200,25){1.5}
\Vertex(200,50){1.5}
\Photon(170,0)(200,25){3}{4}
\Photon(172,0)(202,25){3}{4}
\DashLine(228,0)(198,25){4}
\DashLine(200,25)(230,0){4}
\DashLine(200,50)(170,75){4}
\Photon(230,75)(200,50){3}{4}
\DashLine(200,25)(200,50){4}

\Text(250,35)[]{\large c}
\Vertex(300,25){1.5}
\Photon(270,0)(300,25){3}{4}
\Photon(272,0)(302,25){3}{4}
\DashLine(330,0)(300,25){4}
\DashLine(328,0)(298,25){4}
\Photon(300,25)(330,50){3}{4}
\DashLine(300,25)(280,50){4}

\Text(350,35)[]{\large d}
\Vertex(430,25){1.5}
\Vertex(370,25){1.5}
\Photon(350,0)(370,25){3}{3}
\Photon(352,0)(372,25){3}{3}
\DashLine(450,0)(430,25){4}
\DashLine(448,0)(428,25){4}
\Photon(370,25)(450,50){3}{8}
\DashLine(430,25)(350,50){4}
\Photon(430,26)(370,26){3}{6}
\Photon(430,24)(370,24){3}{6}
\end{picture}

\subsection{Fermion/Higgs Coannihilation}

\subsubsection{$\Phi^{(1)} f^{(1)} \rightarrow \Phi^{(0)} f^{(0)} $}

This coannihilation diagram includes processes to external Goldstone bosons
which are of course just the leading order contribution to the analogous
process of coannihilating into external SM gauge bosons.

\begin{picture}(450,50)(0,0)
\Text(0,35)[]{\large a}
\Vertex(80,25){1.5}
\Vertex(20,25){1.5}
\ArrowLine(0,0)(20,25)
\ArrowLine(02,0)(22,25)
\DashLine(100,0)(80,25){4}
\DashLine(98,0)(78,25){4}
\DashLine(80,25)(100,50){4}
\ArrowLine(20,25)(0,50)
\Photon(80,26)(20,26){3}{6}
\Photon(80,24)(20,24){3}{6}
\end{picture}

\section{Cross Sections} 
\label{CrossSect}

\subsection{Fermion Annihilation}

The following sections refer to the diagrams and matrix elements computed 
in Sec.~\ref{Diagrams}, by the Section numbers assigned to each process.  
The total cross section is the sum of the appropriate terms and cross-terms, 
multiplied by the necessary spin averaging and identical particle factors.
We use the notation $L= \ln\left ( \frac{1+\beta}{1-\beta} \right)$.

\subsubsection{$e_R$}

\begin{center}
\begin{tabular}{|c|c|c|}
\hline
Process & Diagrams & Total cross section \\ 
\hline
 $E_R E_R \rightarrow e_R e_R $ & C.1.1 a,b & $\frac{g_1^4 Y_{e_R}^4}{32 
\pi \beta^2 
s^2 m_{KK}^2} [\beta s(2s -m_{KK}^2)+m_{KK}^2(4s-5m_{KK}^2)L ]$ \\ 
$E_R \overline{E_R} \rightarrow e \overline{e} $ & C.1.1 c,d & $ 
\frac{g_1^4 
Y_{e_R}^4 }{192 \pi \beta^2 s^2} [\beta(12s^2+16 m_{KK}^4+65sm_{KK}^2)-
12m_{KK}^2(4s+3 m_{KK}^2 )L ]  $ \\
& & $+\frac{g_1^4 Y_{e_R}^2 Y_{e_L}^2}{24 
\pi 
\beta s^2} (s+2m_{KK}^2) $\\
$E_R \overline{E_R} \rightarrow f \overline{f} $ & C.1.1 d & $\frac{g_1^4 Y_{e_R}^2
(s+2 m_{KK}^2 )}{24 \pi \beta s^2 }  \sum_{f=\neq e} Nc 
(Y_{f_R}^2+Y_{f_L}^2) $ \\
$E_R \overline{\mu_R} \rightarrow e_R \overline{\mu_R} $ & C.1.1 c & $ \frac{g_1^4 
Y_{e_R}^4}{64 \pi \beta^2 s m_{KK}^2} [\beta (4s+9 m_{KK}^2) -8 
m_{KK}^2L] $\\
$E_R \mu_R \rightarrow e_R \mu_R $ & C.1.1 a & $ \frac{g_1^4 Y_{e_R}^4}{64 
\pi \beta s m_{KK}^2 } (4 s -3 m_{KK}^2 )$ \\
\hline
$ E_R \overline{E_R} \rightarrow \phi \phi^* $ &C.1.3  & $\frac{g_{Ze}^2 
g_{Z\phi}^2}{48 \pi \beta s^2} (s+2 m_{KK}^2 ) $\\ 
\hline
$E_R \overline{E_R} \rightarrow \gamma \gamma $& C.1.2 a,b &$ 
\frac{e^4}{8 \pi 
s^3 \beta^2} \left\{(s^2+4 m_{KK}^2s -8 m_{KK}^4)L -s\beta(s+4 m_{KK}^2) \right\} $\\
$E_R \overline{E_R} \rightarrow Z Z $& C.1.2 a,b & 
$\frac{g^4 \left( \frac{2(-1/2 + \sin (\theta_W))}{sin(2 \theta_W} \right)^4}{
8 \pi s^3 \beta^2} 
\left\{(s^2+4 m_{KK}^2s -8 m_{KK}^4)L -s\beta(s+4 m_{KK}^2) \right\} $\\
$E_R \overline{E_R} \rightarrow W W$ & C.1.2 c & $ 
\,{\frac {e^4(s+2\,{{\it m_{KK}}}^{2})}{192\pi \,{s}^{2}\beta\,{c_
{{w}}}^{4}}}$ \\
\hline
\end{tabular}
\end{center}

Note that in the $s$-channel diagrams the mass of the propagator was
ignored (i.e., only hypercharge gauge photon exchanged), since we
neglected terms of order $v^2/m_{KK}^2$.

\subsubsection{$e_L$}

\begin{center}
\begin{tabular}{|c|c|c|}
\hline
Process & Diagrams & Total cross section \\ 
 \hline $E_L \overline{E_L} \rightarrow e'_L \overline{e'_L} $ & C.1.1 c,d & $ 
\frac{ g_{s}^2 g_{t,e}^2[  5\,s\beta-2\, \left( 2\,s+3\,{m_{{{\it 
KK}}}}^{2}
 \right) L]}{32 \pi s^2 \beta^2}+{\frac{g_{t,e}^4 [\beta\, \left( 
4\,s+9\,{
m_{{{\it KK}}}}^{2} \right) -8\,{m_{{{\it KK}}}}^{2}L]}{64\pi s \beta^2 
m_{\it KK}^{2}}} $\\
& & $+{\frac {g_{s}^4  
[s+2\,{m_{{{\it KK}}}}^{2}]}{24\pi s^2\beta}}$\\
\hline
$E_L \overline{E_L} \rightarrow W^+ W^- $ & C.1.2 a,c & $
\,g^4 {\frac { \left( 96\,{c_{{w}}}^{4}{m_{{{\it KK}}}}^{2
}+24\,s{c_{{w}}}^{4} \right) L+2\,\beta\,{m_{{{\it KK}}}}^{2}+s\beta-
40\,s{c_{{w}}}^{4}\beta-176\,\beta\,{m_{{{\it KK}}}}^{2}{c_{{w}}}^{4}}
{768\pi \,{s}^{2}{\beta}^{2}{c_{{w}}}^{4}}} $ \\
\hline
\end{tabular}
\end{center}

Cross sections identical in nature to those of $E_R$ have not been 
reproduced in this table.  They can be obtained from the previous table 
with appropriate replacements of the coupling constants.

In the cross section to leptons of the same generation, explicit coupling 
factors have been kept since the diagrams are the same for both (i) $E_L 
\overline{E_L} \rightarrow e_L \overline{e_L}$ and (ii) $E_L \overline{E_L} 
\rightarrow \nu_e \overline{\nu_e}$, 
as well as the analogous processes with neutrinos in the 
initial state (iii) and (iv).  Here $g^2_{s (t)}$ is the total coupling 
in the $s$-($t$-)channel.
Ignoring the SM gauge boson masses, 
$g_s^2 =g_t^2 = g^2 (\tan^2 \theta_W Y_{E_L} Y_{e_L} + T_3 (E_L) T_3 (e_L))$
for annihilation into the same particle, and 
$g_s^2 = g^2 (\tan^2 \theta_W Y_{E_L} Y_{\nu_L} + T_3 (E_L) T_3 (\nu_L))$
$g_t^2 = g^2/2$ for processes to the electroweak partner (where the 
$t$-channel is mediated by a $W^{\pm}$ boson).

For cross sections to gauge bosons, the leading terms in the limit of
vanishing gauge boson mass have been kept.

\subsubsection{quarks}

The processes 
$QQ \rightarrow qq$ and $Q \overline{Q} \rightarrow q \overline{q}$ 
are essentially the same as the corresponding processes for electrons 
shown above, except that an additional diagram involving gluon exchange 
is present.  In the limit that all KK gauge bosons have mass, and we 
ignore the masses of all SM gauge bosons in the $s$ channel, the relevant 
amplitude comes from the electron amplitude (with the appropriate 
modification of the hypercharge value) 
plus the same amplitude with all couplings changed to $g_s$.  
Annihilation to scalars and electroweak  
gauge bosons are identical to the electron case, with the appropriate 
couplings.  Thus the only novel diagrams are annihilation to $gg$ and 
$g \gamma$.  

\begin{center}
\begin{tabular}{|c|c|c|}
\hline
 Process & Diagrams & Total cross section \\ 
\hline
$Q \overline{Q} \rightarrow g \gamma $& C.1.2 a,b & 
$\frac{e^2 g_s^2}{2}\,{\frac { 
\left( -8\,{m_{{{\it KK}}}}^{4}+4\,{m_{{{\it KK}}}}^{2}s
+{s}^{2} \right) L-{s}^{2}\beta-4\,\beta\,{m_{{{\it KK}}}}^{2}s}{18\pi 
\,{s}^{3}{\beta}^{2}}} $\\
$Q \overline {Q} \rightarrow g Z $& C.1.2 a & 
$\frac{g_s^2g_{Z,q}^2}{36}\,{\frac { \left( {s}^{2}+4\,s{m_{{{\it 
KK}}}}^{2}-8\,{m_{{{\it KK
}}}}^{4} \right) L-{s}^{2}\beta-4\,s\beta\,{m_{{{\it KK}}}}^{2}}{{
\beta}^{2}{s}^{3}\pi }}$ \\
$Q \overline Q \rightarrow g g$ & C.1.2 a,b,c &$ 
-\frac{g_s^4}{216}{\frac{-4\,\left 
({m_{KK}}^{4}+4\,s{m_{KK}}^{2}+
\,{s}^{2}\right ) L+7\,\beta\,{s}^{2}+31\,s\beta\,{m_{KK}}^{2}}{ \pi 
{s}^{3} {\beta}^{2}}}$\\
\hline
\end{tabular}
\end{center}

Here $g_{Z,q}$ is the coupling of the quark to the Z boson.
Note that in annihilation to gluons, the appropriate color factors 
have been included, along with a factor of $\frac{1}{9}$ for the average 
over quark colors.  In $qq \rightarrow g \gamma$ or $qq \rightarrow g Z$, a 
factor of $1/2$ must be included for the trace over $t^a t^b$.

\subsection{Gauge Boson Annihilation}

Because the KK Weinberg angle is sufficiently small that its effect on 
our computations can be neglected, the results here are presented in the
$B^{(1)}$,  $W^{3(1)}$ basis.

\subsubsection{Hypercharge}

\begin{center}
\begin{tabular}{|c|c|c|}
\hline
Process & Diagrams & Total cross section \\ 
\hline
$B_{\mu} B_{\nu} \rightarrow f \overline{f} $& C.2.1 a,b &$ 
\,{\frac {g_1^4 Y_f^4 [(10\,{m_{{{\it KK}}}}^{2}
+5\,s)L-7\,s\beta]}{72
{s}^{2}\pi \,{\beta}^{2}}}$ \\
$B_{\mu} B_{\nu} \rightarrow W^+ W^- $ & C.2.3 d,e,f & 
${\frac {1}{192}}\,{\frac {{{\it s}_{{w}}}^{4}{g}^{4}}{\beta\,{c_{{w}}
}^{4}\pi \,s}} $\\
$B_{\mu} B_{\nu} \rightarrow Z Z $ & C.2.3 d,e,f & 
$\frac{1}{2}{\frac {1}{192}}\,{\frac {{{\it 
s}_{{w}}}^{4}{g}^{4}}{\beta\,{c_{{w}}
}^{4}\pi \,s}} $\\
$B_{\mu} B_{\nu} \rightarrow h h $ & C.2.2 a,b,c & 
$\frac{1}{2}{\frac {1}{192}}\,{\frac {{{\it 
s}_{{w}}}^{4}{g}^{4}}{\beta\,{c_{{w}}
}^{4}\pi \,s}} $\\
\hline
\end{tabular}
\end{center}

\subsection{$W_3$}

\begin{center}
\begin{tabular}{|c|c|c|}
\hline
Process & Diagrams & Total cross section \\ 
\hline
$W_{3\mu} W_{3\nu} \rightarrow f \overline{f} $& C.2.1 a,b & $ 
\,{\frac {g^4 T_3(f)^4 [(10\,{m_{{{\it KK}}}}^{2}
+5\,s)L-7\,s\beta]}{72{s}^{2}\pi \,{\beta}^{2}}}$ \\
$W_{3\mu} W_{3\nu} \rightarrow \phi \phi $& C.2.2 a,b,c & 
${\frac{1}{2}\frac {g^4 }{192 \pi \,s\beta}} $ \\
$W_{3\mu} W_{3\nu} \rightarrow Z_{\mu}Z_{\nu} $ & C.2.3 d,e,f & 
$\frac{1}{2} 
{\frac {g^4}{192\beta\,\pi \,s}}$ \\
$W_{3\mu} W_{3\nu} 
\rightarrow W^+ W^- $ & C.2.3 a,b,c,d,e,f & $ 
\frac{{g}^4}{18}\,{\frac { 
\left( -12\,{m_{{{\it KK}}}}^{4}s+24\,{m_{{{\it KK}}}}^
{6} \right) L+4\,\beta\,{s}^{3}+3\,\beta\,{m_{{{\it KK}}}}^{2}{s}^{2}+
12\,s{m_{{{\it KK}}}}^{4}\beta}{\pi \,{s}^{3}{\beta}^{2}{m_{{{\it KK}}
}}^{2}}}$ \\
& &$ + {\frac {1}{192}}\,{\frac{{g}^{4}}{\beta\,\pi \,s}} $ \\
\hline
\end{tabular}
\end{center}

\subsection{$W^\pm$}

\begin{center}
\begin{tabular}{|c|c|c|}
\hline
Process & Diagrams & Total cross section \\ 
\hline
$W^+_{\mu} W^-_{\nu} \rightarrow f_R \overline{f_R} $& C.2.1 c & 0 to leading 
order in the SM z boson mass \\
$W^+_{\mu} W^-_{\nu} \rightarrow f_L \overline{f_L} $& C.2.1 a,c & $ {\frac 
{g^4}{8}}\,{\frac { \left( 5\,s+12\,{m_{{{\it KK}}}}^{2}
 \right) L-10\,\beta\,s-8\,\beta\,{m_{{{\it KK}}}}^{2}}{72 \pi \,{s}^{2}{
\beta}^{2}}}$ \\
$W^+_{\mu} W^-_{\nu} \rightarrow W^+_\nu W^-_\mu $ & C.2.3 a,c,d,e,s & $ 
\,\frac{g^4}{144}{\frac { \left( -32\,{m_{{{\it KK}}}}^{2}s-48\,{m_{{
{\it KK}}}}^{4} \right) L+16\,{s}^{2}\beta+41\,{m_{{{\it KK}}}}^{2}
\beta\,s+87\,{m_{{{\it KK}}}}^{4}\beta}{\pi \,{s}^{2}{\beta}^{2}{m_{{{
\it KK}}}}^{2}}}$ \\
$W^+_{\mu} W^-_{\nu} \rightarrow A_{\nu}A_{\mu} $ & C.2.3 a,b,c & 
$\,\frac{e^4}{2}{\frac { \left( -12\,{m_{{{\it KK}}}}^{4}s+24\,{m_{{{\it 
KK}}}}^
{6} \right) L+4\,\beta\,{s}^{3}+3\,\beta\,{m_{{{\it KK}}}}^{2}{s}^{2}+
12\,{m_{{{\it KK}}}}^{4}s\beta}{18 \pi \,{\beta}^{2}{m_{{{\it KK}}}}^{2}{
s}^{3}}}$ \\
 $W^+_{\mu} W^-_{\nu} \rightarrow Z_{\nu} Z_{\mu} $ & C.2.3 a,b,c,d,e,f &
$ \,{\frac {{g^4c_{{w}}}^{4} \left( 24\,L{m_{{{\it 
KK}}}}^{6}+12\,{m_{{
{\it KK}}}}^{4}s\beta-12\,L{m_{{{\it KK}}}}^{4}s+3\,{s}^{2}\beta\,{m_{
{{\it KK}}}}^{2}+4\,{s}^{3}\beta \right) }{36\pi \,{s}^{3}{\beta}^{2}{m_
{{{\it KK}}}}^{2}}}$ \\
&& $+{\frac {1}{384}}\,{\frac {{g}^{4}}{\beta\,\pi 
\,s}}$ \\  
$W^+_{\mu} W^-_{\nu} \rightarrow Z_{\nu} A_{\mu} $ & C.2.3 a,b,c & 
$\frac{e^2 c_w^2 g^2}{18}{\frac { \left( -12\,{m_{{{\it 
KK}}}}^{4}s+24\,{m_{{{\it KK}}}}^
{6} \right) L+4\,\beta\,{s}^{3}+3\,\beta\,{m_{{{\it KK}}}}^{2}{s}^{2}+
12\,s\beta\,{m_{{{\it KK}}}}^{4}}{\pi \,{s}^{3}{\beta}^{2}{m_{{{\it KK
}}}}^{2}}}$ \\
$W^+_{\mu} W^-_{\nu} \rightarrow h h $ & 2.2.3 a,b,c & 
${\frac{1}{2}\frac {g^4 }{192 \pi \,s\beta}} $\\
 $W^+_{\mu} W^-_{\nu} \rightarrow h z $ & 2.2.3 a,b,d & 
${\frac {g^4}{576}}\,{\frac {s-4\,{m_{{{\it KK}}}}^{2}}{\pi \,{s}^{2}\beta
}}$ \\  
\hline
\end{tabular}
\end{center}

\subsection{Gluon}

\begin{center}
\begin{tabular}{|c|c|c|}
\hline
Process & Diagrams & Total cross section \\ 
\hline
$g_{\mu} g_{\nu} \rightarrow f \overline{f} $& C.2.1 a,b,c & ${\frac 
{g_s^4}{64}}\,{\frac { \left( 40\,s+98\,{m_{{{\it KK}}}}^{2}
 \right) L-83\,\beta\,s-72\,\beta\,{m_{{{\it KK}}}}^{2}}{6\pi \,{s}^{2}
{\beta}^{2}}}$ \\  
$g_{\mu} g_{\nu} \rightarrow g_{\nu} g_{\mu} $ & C.2.3 a,b,c,s & 
$\,{g_s^4\frac { \left( -8\,{m_{{{\it KK}}}}^{2}{s}^{2}-24\,{m_{{{\it KK
}}}}^{4}s+24\,{m_{{{\it KK}}}}^{6} \right) L+8\,{s}^{3}\beta+13\,{m_{{
{\it KK}}}}^{2}\beta\,{s}^{2}+34\,{m_{{{\it KK}}}}^{4}s\beta}{32\pi \,{s
}^{3}{\beta}^{2}{m_{{{\it KK}}}}^{2}}}$ \\
\hline
\end{tabular}
\end{center}

Here the factor of $1/64$ was included for the average over 
initial gluon colors.

\subsection{Scalar Annihilation}

In this set of cross sections, note that the KK modes of the 
SM Goldstone bosons are in fact separate propagating degrees of freedom, 
and thus their annihilation cross sections must be calculated 
independently of those of the KK W and Z bosons.  For processes 
to one Higgs and one SM gauge bosons, we use the Goldstone boson 
approximation, as processes to final state transverse polarizations 
are suppressed by $v/m_{KK}$.

\begin{center}
\begin{tabular}{|c|c|c|}
\multicolumn{3}{l}{Scalar Annihilation into Scalars:} \\
\hline
Process & Diagrams & Total cross section \\ 
\hline
$ w^+ w^- \rightarrow h h$ & C.3.1 a,b,d & $
\,{\frac {{H}^{2}}{512\pi \,s\beta\,{c_{{w}}}^{4}}}+
\,{\frac { \left( 4\,\beta\,{m_{{{\it KK}}}}^{2}-8\,{m_
{{{\it KK}}}}^{2}L \right) H}{512\pi \,{\beta}^{2}s{m_{{{\it KK}}}}^{2}{c
_{{w}}}^{2}}}+\,{\frac {8\,\beta\,s+4\,\beta\,{m_{{{
\it KK}}}}^{2}}{512\pi \,{\beta}^{2}s{m_{{{\it KK}}}}^{2}}} 
$ \\
$ w^+ w^- \rightarrow z h$ & C.3.1 a,b,c & $
{\frac {1}{768}}\,{\frac {24\,{s}^{2}\beta-92\,\beta\,s{m_{{{\it KK}}}
}^{2}-16\,\beta\,{m_{{{\it KK}}}}^{4}}{\pi \,{\beta}^{2}{s}^{2}{m_{{{
\it KK}}}}^{2}}} $ \\
& &$ +{\frac {1}{768}}\,{\frac {-24\,{m_{{{\it KK}}}}^{2}sL
+44\,\beta\,s{m_{{{\it KK}}}}^{2}+16\,\beta\,{m_{{{\it KK}}}}^{4}}{
\pi \,{\beta}^{2}{s}^{2}{m_{{{\it KK}}}}^{2}{c_{{w}}}^{2}}}+{\frac {1}
{768}}\,{\frac {-4\,\beta\,{m_{{{\it KK}}}}^{4}+\beta\,s{m_{{{\it KK}}
}}^{2}}{\pi \,{\beta}^{2}{s}^{2}{m_{{{\it KK}}}}^{2}{c_{{w}}}^{4}}} $ \\
$z z \rightarrow h h $ & 
C.3.1 a,b,d & ${\frac {1}{512}}\,{\frac {8\,s\beta+4\,\beta\,{m_{{{\it 
KK}}}}^{2}-4\,
\beta\,{m_{{{\it KK}}}}^{2}H^2+8\,{m_{{{\it KK}}}}^{2}H^2 
L+H^{4}\beta\,{m_{{{\it KK}}}}^{2}}{\pi \,{
\beta}^{2}s{m_{{{\it KK}}}}^{2}{c_{{w}}}^{4}}}$ \\
 $hh \rightarrow hh $ & 2.2.3 d  & ${\frac 
{g^4}{512}}\,{\frac {{H}^{4}}{\pi \,s\beta\,{c_{{w}}
}^{4}}} $\\ 
$h h \rightarrow z h $ & 
 & $\approx\left (\frac{m_Z}{M_{KK}}\right)^4  $\\
$h z \rightarrow h h $ & 
 & $\approx\left (\frac{m_Z}{M_{KK}}\right)^4  $\\
 $w^+ z \rightarrow w^+ h $ & C.3.1 a,b,c & $
{\frac {1}{768}}\,{\frac {{g}^{4} \left( 76\,s{m_{{{\it KK}}}}^{2}
\beta-4\,\beta\,{m_{{{\it KK}}}}^{4}+60\,{s}^{2}\beta-144\,s{m_{{{\it 
KK}}}}^{2}L \right) }{\pi \,{\beta}^{2}{s}^{2}{m_{{{\it KK}}}}^{2}}} $\\
&& $+{
\frac {1}{768}}\,{\frac {{g}^{4} \left( 84\,s{m_{{{\it KK}}}}^{2}L-48
\,{s}^{2}\beta-30\,s{m_{{{\it KK}}}}^{2}\beta \right) }{\pi \,{\beta}^
{2}{s}^{2}{m_{{{\it KK}}}}^{2}{c_{{w}}}^{2}}}$ \\
&& $+{\frac {1}{768}}\,{
\frac {{g}^{4} \left( 3\,s{m_{{{\it KK}}}}^{2}\beta+12\,{s}^{2}\beta-
12\,s{m_{{{\it KK}}}}^{2}L \right) }{\pi \,{\beta}^{2}{s}^{2}{m_{{{
\it KK}}}}^{2}{c_{{w}}}^{4}}}$ \\
$ w h \rightarrow w h $ & C.3.1  b,c,d &  $ {\frac {g^4}{768}}\,{\frac 
{12\,{s}^{2}\beta-20\,\beta\,s{m_{{{\it KK}}}
}^{2}-4\,\beta\,{m_{{{\it KK}}}}^{4}}{\pi \,{\beta}^{2}{s}^{2}{m_{{{
\it KK}}}}^{2}}}$ \\
&& $ +{\frac {g^4}{768}}\,{\frac {\left (12\,s{m_{{{\it 
KK}}}}^{2}L-6\,\beta\,s{m_{{{\it KK}}}}^{2}\right ) H^2}{\pi \,{
\beta}^{2}{s}^{2}{m_{{{\it KK}}}}^{2}{c_{{w}}}^{2}}}+{\frac {g^4}{256}}
\,{\frac {{H}^{4}}{\pi \,s\beta\,{c_{{w}}}^{4}}}$ \\ 
$h z \rightarrow z h 
$ & C.3.1 a,c,d & 
$ {\frac {g^4}{768}}
\,{\frac { \left( -24\,s{m_{{{\it KK}}}}^{2}+12\,s{m_{
{{\it KK}}}}^{2}H^2 \right) L}{\pi \,{\beta}^{2}{s}^{2}{m_{{{\it 
KK}}}}^{2}{c_{{w}}}^{4}}}$ \\
& & $ +{\frac {g^4}{768}}\frac{3\,H^{4}\beta
\,s{m_{{{\it KK}}}}^{2}-6\,s{m_{{{\it KK}}}}^{2}\beta\,H^2-
4\,\beta\,{m_{{{\it KK}}}}^{4}+28\,s{m_{{{\it KK}}}}^{2}\beta+12\,{s}^
{2}\beta}{\pi \,{\beta}^{2}{s}^{2}{m_{{{\it KK}}}}^{2}{c_{{w}}}^{4}}$ \\
\hline
\end{tabular} 
\end{center}

\begin{center}
\begin{tabular}{|c|c|c|}
\multicolumn{3}{l}{Scalar Annihilation into Gauge Bosons:} \\
\hline
Process & Diagrams & Total cross section \\ 
\hline
$z z \rightarrow Z Z $& C.3.2 a,b,c,g & $\frac{g^4}{64}{\frac { \left( 
-4\,s{m_{{{\it 
KK}}}}^{2}+8\,{m_{{{\it KK}}}}^{4}
 \right) L+4\,\beta\,s{m_{{{\it KK}}}}^{2}+{s}^{2}\beta}{ \pi  
c_w^4\,{s}^{3}{\beta}^{2}}}+ \,{\frac { g^4{H}^{4}}{512 \pi \,s\beta
\,{c_{{w}}}^{4}}}
$ \\
$z z \rightarrow W^+ W^- $& C.3.2 a,b,c,e,f,g & $
-\,{\frac { \left( -16\,{m_{{{\it KK}}}}^{6}+8\,{m_{{{
\it KK}}}}^{4}s \right) L
-2\,\beta\,{s}^{3}-3\,\beta\,{s}^{2}{m_{{
{\it KK}}}}^{2}-8\,\beta\,s{m_{{{\it KK}}}}^{4}
}{64 \pi \,{s}^{3}{\beta}^{2}{m_{{{\it KK}}}}^{2}}}$ \\
& & $+
\frac {g^4 H^2}{64}\,\frac {\beta-2\,L}{c_{w}^{2}s\pi \,{\beta}^{2}
}+ {\frac {g^4}{256}}\,{\frac {H^4}{\beta\,\pi \,s{c_{{w}}}^{4}}}$ \\
$ h h \rightarrow Z Z $ & C.3.2 a,b,c,e,f,g & $ 
{\frac {g^4}{64}}\,{\frac { \left( -4\,s{{\it Mh}}^{2}+8\,{{\it Mh}}^{4}
 \right) L
+{s}^{2}\beta+4\,{{\it Mh}}^{2}\beta\,s
}{\pi \,{s}^{3}{\beta}^{2}{c}^{4}}}
+{\frac {1}{512}}\,{\frac {{H}^{4}}{\beta\,\pi \,s{c_{{w}}}^{4}}}$ \\ 
& & $
{\frac {1}{512}}\,{\frac { \left( -4\,\beta\,{m_{{{\it KK}}}}^{2}+8\,{
m_{{{\it KK}}}}^{2}L \right) {H}^{2}}{\pi \,{\beta}^{2}s{m_{{{\it KK}}
}}^{2}{c_{{w}}}^{4}}}+{\frac {1}{512}}\,{\frac {8\,\beta\,s+4\,\beta\,
{m_{{{\it KK}}}}^{2}}{\pi \,{\beta}^{2}s{m_{{{\it KK}}}}^{2}{c_{{w}}}^
{4}}}
$ \\
$ hh \rightarrow W^+ W^- $ & C.3.2 a,b,c,e,f,g & 
$\frac{g^4}{32}\left (\,{\frac { \left( -4\,s{{\it Mh}}^{2}+8\,{{\it 
Mh}}^{4} 
\right) L
+{s}^{2}\beta+4\,{{\it Mh}}^{
2}\beta\,s}{\pi \,{s}^{3}{\beta}^{2}}}\right )+{\frac {g^4}{256}}\,{\frac 
{{H}^{
4}}{\beta\,\pi \,s{c_{{w}}}^{4}}}$ \\
& & $+{\frac {g^4}{256}}\,{\frac { \left( 4
\,{m_{{{\it KK}}}}^{2}{c_{{w}}}^{2}\beta-8\,{m_{{{\it KK}}}}^{2}{c_{{w
}}}^{2}L \right) {H}^{2}}{\pi \,{\beta}^{2}s{m_{{{\it KK}}}}^{2}{c_{{w
}}}^{4}}}+{\frac {g^4}{256}}\,{\frac {8\,{c_{{w}}}^{4}\beta\,s+4\,{c_{{w
}}}^{4}\beta\,{m_{{{\it KK}}}}^{2}}{\pi \,{\beta}^{2}s{m_{{{\it KK}}}}
^{2}{c_{{w}}}^{4}}}$ \\
&& \\
$z \phi \rightarrow W^+ W^- $& C.3.2 a,b,d,e,f & $
{\frac {g^4}{384}}\,{\frac {5\,{s}^{2}\beta-48\,s{m_{{{
\it KK}}}}^{2}L+124\,s{m_{{{\it KK}}}}^{2}\beta-96\,L{m_{{{\it KK}}}}^
{4}}{\pi \,{s}^{3}{\beta}^{2}}}$ \\
& & $+{\frac {g^4}{768}}
{\frac {-24 m_{KK}^2 s L+
24\,{s}^{2}\beta-47\,s{m_{{{\it KK}}}}^{2}
\beta-4\,\beta\,{m_{{{\it KK}}}}^{4}}{\pi \,{s}^{2}{\beta}^{2}{m_{{{
\it KK}}}}^{2}}} $ \\
$w^+ w^- \rightarrow W^+ W^- $& C.3.2 a,c,d,e,f,g & $ 
\frac {g^4}{384} \, {\frac {17\,\beta\,s-96\,{m_{{{\it KK}}}}^{2}L+172\,
\beta\,{m_{{{\it KK}}}}^{2}}{\pi \,{s}^{2}{\beta}^{2}}}+{\frac {g^4}{64}
}\,{\frac {{H}^{4}}{\beta\,\pi \,s{c_{{w}}}^{4}}}$ \\
& & $ \frac{g^4}{192} \left[
\frac{\left( -3\,s{m_{{{\it KK}}}}^{2}\beta+6\,s{m_{{{\it KK}}}}^{2}
L \right) {H}^{2}}{\pi \,{s}^{2}{\beta}^{2}{m_{{{\it KK}}}}^{2}{c_{{w}}}^{4}}
+\frac{3\,{s}^{2}\beta+7\,s{m_{{{\it KK}}}}^{2}\beta
-6\,s{m_{{{\it KK}}}}^{2}L-\beta\,{m_{{{\it KK}}}}^{4}}
{\pi \,{s}^{2}{\beta}^{2}{m_{{{\it KK}}}}^{2}{c_{{w}}}^{4}} \right]$ \\
$w^+ w^- \rightarrow Z Z  $& C.3.2 a,b,c,e,f,g & $ {\frac {1}{64}}\,{\frac { 
g^4 \left( 2\,{c_{{w}}}^{2}-1 \right) ^{4}
 \left( 8\,{m_{{{\it KK}}}}^{4}L-4\,{m_{{{\it KK}}}}^{2}Ls+{s}^{2}
\beta+4\,s{m_{{{\it KK}}}}^{2}\beta \right) }{\pi \,{s}^{3}{\beta}^{2}
{c_{{w}}}^{4}}}$ \\
& & $
\,{\frac {g^4{ H}^{4}}{512 s\beta\,\pi \,{c_{{w}}}^{4}}}+ {\frac {g^4 
\left( 4\,\beta\,{m_{{{\it KK}}}}^{2}{c_{{w}
}}^{2}-8\,{m_{{{\it KK}}}}^{2}L{c_{{w}}}^{2} \right) {H}^{2}}{512{m_{{{
\it KK}}}}^{2}\pi \,s{\beta}^{2}{c_{{w}}}^{4}}}$ \\
&&$+\,{
\frac {8\,{c_{{w}}}^{4}s\beta+4\,\beta\,{m_{{{\it KK}}}}^{2}{c_{{w}}}^
{4}}{512{m_{{{\it KK}}}}^{2}\pi \,s{\beta}^{2}{c_{{w}}}^{4}}} $ \\
$w^+ w^- \rightarrow A A  $& C.3.2 a,b, c  & $ e^4 \,{\frac { \left( 
8\,{m_{{{\it KK}}}}^{4}-4\,s{m_{{{\it KK}}}}^{2}
 \right) L+4\,\beta\,s{m_{{{\it KK}}}}^{2}+{s}^{2}\beta}{2\pi \,{s}^{3}
{\beta}^{2}}}
$ \\
$w^+ w^- \rightarrow A Z  $& C.3.2 a,b,c  & $\frac{g^4 
s_{w}^2(2c_w^2-1)^2}{c_w^2}\,{\frac { \left( 
-4\,s{m_{{{\it KK}}}}^{2}+8\,{m_{{{\it KK}}}}^{4}
 \right) L+4\,\beta\,s{m_{{{\it KK}}}}^{2}+{s}^{2}\beta}{8\pi \,{s}^{3}
{\beta}^{2}}}
$ \\
$w^+ h \rightarrow W^{+} A  $& C.3.2 a,b,c,d & $
-{\frac {1}{384}}\,{\frac {{{\it s}_{{w}}}^{2}{g}^{4} \left( 17\,
\beta\,s-96\,{m_{{{\it KK}}}}^{2}L+172\,\beta\,{m_{{{\it KK}}}}^{2}
 \right) }{\pi \,{s}^{2}{\beta}^{2}}} $ \\
$w^+ z  \rightarrow W^{+} A  $& C.3.2 a,b, c,d & $        
-{\frac {1}{384}}\,{\frac {{{\it s}_{{w}}}^{2}{g}^{4} \left( 17\,
\beta\,s-96\,{m_{{{\it KK}}}}^{2}L+172\,\beta\,{m_{{{\it KK}}}}^{2}
 \right) }{\pi \,{s}^{2}{\beta}^{2}}}$ \\
$w^+ h  \rightarrow W^{+} Z$ & C.3.2 a,b,c,d,e,f,g & 
[Eq.~(\ref{overflow1}) below] \\
$w^+ z  \rightarrow W^{+} Z$ & C.3.2 a,b,c,d,e,f,g & 
[Eq.~(\ref{overflow2}) below] \\
\hline
\end{tabular}
\end{center}

\begin{eqnarray}
-\frac{g^4}{384} \bigg[ {\frac{-24\,\beta\,{c_{{w}}}^{2}{s}
^{2}+17\,{s}^{2}\beta\,{c_{{w}}}^{4}+12\,{s}^{2}\beta-96\,s{c_{{w}}}^{
4}{m_{{{\it KK}}}}^{2}L+96\,{c_{{w}}}^{2}s{m_{{{\it KK}}}}^{2}L+48\,s{
m_{{{\it KK}}}}^{2}\beta}{\pi \,{s}^{3}{\beta}^{2}{c_{{w}}}^{2}}} \nonumber \\
+ \frac{172\,{c_{{w}}}^{4}s{m_{{{\it KK}}}}^{2}\beta-
48\,s{m_{{{\it KK}}}}^{2}L-96\,{c_{{w}}}^{2}s{m_{{{\it KK}}}}^{2}\beta
-192\,{m_{{{\it KK}}}}^{4}L{c_{{w}}}^{2}+96\,L{m_{{{\it KK}}}}^{4}}{
\pi \,{s}^{3}{\beta}^{2}{c_{{w}}}^{2}} \nonumber \\
- \frac{60\,
{s}^{2}\beta\,{c_{{w}}}^{4}-144\,s{c_{{w}}}^{4}{m_{{{\it KK}}}}^{2}L+
76\,{c_{{w}}}^{4}s{m_{{{\it KK}}}}^{2}\beta-4\,{c_{{w}}}^{4}\beta\,{m_
{{{\it KK}}}}^{4}-48\,\beta\,{c_{{w}}}^{2}{s}^{2}}{
2 \pi \,{s}^{2}{\beta}^{2}{m_{{{\it KK}}}}^{2}{c_{{w}}}^{4}} \nonumber \\
- \frac{-30\,{c_{{w}}}^{2}s{m
_{{{\it KK}}}}^{2}\beta+84\,{c_{{w}}}^{2}s{m_{{{\it KK}}}}^{2}L-12\,s{
m_{{{\it KK}}}}^{2}L+3\,s{m_{{{\it KK}}}}^{2}\beta+12\,{s}^{2}\beta}{
2 \pi \,{s}^{2}{\beta}^{2}{m_{{{\it KK}}}}^{2}{c_{{w}}}^{4}} \bigg] 
\label{overflow1}
\end{eqnarray}

\begin{eqnarray}
- \frac{g^4}{384} \bigg[
\frac{-24\,\beta\,{c_{{w}}}^{2}{s}^
{2}+17\,{s}^{2}\beta\,{c_{{w}}}^{4}+12\,{s}^{2}\beta-96\,s{c_{{w}}}^{4
}{m_{{{\it KK}}}}^{2}L+96\,{c_{{w}}}^{2}s{m_{{{\it KK}}}}^{2}L+48\,s{m
_{{{\it KK}}}}^{2}\beta}{
\pi \,{s}^{3}{\beta}^{2}{c_{{w}}}^{2}} \nonumber \\
+ \frac{172\,{c_{{w}}}^{4}s{m_{{{\it KK}}}}^{2}\beta-
48\,s{m_{{{\it KK}}}}^{2}L-96\,{c_{{w}}}^{2}s{m_{{{\it KK}}}}^{2}\beta
-192\,{m_{{{\it KK}}}}^{4}L{c_{{w}}}^{2}+96\,L{m_{{{\it KK}}}}^{4}}{
\pi \,{s}^{3}{\beta}^{2}{c_{{w}}}^{2}} \nonumber \\
- 
\frac{12\,{s}^{2}\beta-20\,s{m_{{{\it KK}}}}^{2}\beta-
4\,\beta\,{m_{{{\it KK}}}}^{4}}
{2 \pi \,{s}^{2}{\beta}^{2}{m_{{{\it KK}}}}^{2}} 
- 
\frac{12\,{H}^{2}s{m_{{{\it KK}}}}^{2}L-6\,s{m_{{{\it KK}}}}^{2}
\beta\,{H}^{2}}{2 \pi \,{s}^{2}{\beta}^{2}{m_{{{\it KK}}}}^{2}{c_{{w}}}^{2}}
- 
\frac{3 {H}^{4}}{2 \beta\,\pi \,s{c_{{w}}}^{4}} \bigg]
\label{overflow2}
\end{eqnarray}

\begin{center}
\begin{tabular}{|c|c|c|}
\multicolumn{3}{l}{Scalar Annihilation into Fermions:} \\
\hline
Process & Diagrams & Total cross section \\ 
\hline
$w^+ w^- \rightarrow f \overline{f} $  & C.3.3& $\left (\frac{g^2(T_3-Q 
s_{w}^2) (2 c_w^2-1)}{c_w^2}+ e^2 Q \right )^2 \,{\frac 
{s-4\,{m_{{{\it KK}}}}^{2}}{12\pi \,\beta\,{s}^{2}}}
$ \\
$w^+ z (\mbox{or} \phi) \rightarrow f^+ \overline{f^-} $  & C.3.3& $ \left 
(\frac{g^2}{2}\right)^2 \,{\frac 
{s-4\,{m_{{{\it KK}}}}^{2}}{12\pi \,\beta\,{s}^{2}}}
$ \\
$z \phi \rightarrow f \overline{f} $  & C.3.3& $\left (\frac{g^2(T_3-Q s_{w}^2) 
}{2 c_w^2} \right)^2\,{\frac {s-4\,{m_{{{\it KK}}}}^{2}}{12\pi 
\,\beta\,{s}^{2}}} $ \\
\hline
\end{tabular}
\end{center}
  
Here $Q$ and $T_3$ denote the charge and SU(2) charge, respectively, of 
the fermion.

\subsection{Coannihilation}

\subsubsection{Gauge Boson Coannihilation}

\begin{center}
\begin{tabular}{|c|c|c|}
\hline
Process & Diagrams & Total cross section \\ 
\hline
$g_{\mu} A_{\nu}, Z_{\nu} \rightarrow q \overline{q} $& C.2.1 a,b &$ 
\,\frac{1}{2(N_c^2-1)} {\frac 
{(g_{Zq}^2 g_s^2) [(10\,{m_{{{\it KK}}}}^{2} +5\,s)L-7\,s\beta]}{72
{s}^{2}\pi \,{\beta}^{2}}}$ \\  
$g_{\mu} W^{+}_{\nu} \rightarrow q \overline{q} $& C.2.1 a,b &$\,\frac{1}{2(N_c^2-1)} {\frac 
{(g_{Wq}^2 g_s^2) [(10\,{m_{{{\it KK}}}}^{2} +5\,s)L-7\,s\beta]}{72
{s}^{2}\pi \,{\beta}^{2}}}$ \\    
$W^\pm_{\mu}  W^{(3)}_{\nu} \rightarrow f_L \overline{f_L} $& C.2.1 a,b,c 
&$
{\frac {g^4}{576}}\,{\frac { \left( 5\,s+14\,{m_{{{\it KK}}}}^{2}
 \right) L+ \left( -16\,{m_{{{\it KK}}}}^{2}-13\,s \right) \beta}{{s}^
{2}\pi \,{\beta}^{2}}}
$ \\
$W^{\pm}_{\mu}  B_{\nu} \rightarrow f \overline{f} $& C.2.1 a,b &${\frac 
{-1}{144}}\,{\frac {Y_f^2 g_1^2g^2 \left(  \left( -10\,{m_{{{\it 
KK}}}}^{2}-5
\,s \right) L+7\,s\beta \right)}{{s}^{2}\pi \,{\beta}^{2}}} $ \\
$B_{\mu}  W^{(3)}_{\nu} \rightarrow f \overline{f} $& C.2.1 a,b & as for $ BB 
\rightarrow ff$ \\ 
$W^{\pm}_{\mu} W^{(3)}_{\nu} \rightarrow W^\pm_{\nu} A_{\mu} $ & C.2.3  
a,c,s & $ 
\frac{g^2 e^2}{18}{\frac { \left( -4\,{m_{{{\it KK}}}}^{2}s-6\,{m_{{{\it 
KK}}}}^{4
} \right) L+2\,{s}^{2}\beta+5\,s\beta\,{m_{{{\it KK}}}}^{2}+11\,\beta
\,{m_{{{\it KK}}}}^{4}}{ {m_{{{\it KK}}}}^{2}\pi \,{s}^{2}{\beta}^{2}}}$ \\  
$W^\pm_{\mu} W^{(3)}_{ \nu} \rightarrow W^\pm_{\nu} Z_{\mu} $ & C.2.3  
a,c,s,e,d& $ 
\frac{g^4 c_w^2}{18}{\frac { \left( -4\,{m_{{{\it KK}}}}^{2}s-6\,{m_{{{\it 
KK}}}}^{4
} \right) L+2\,{s}^{2}\beta+5\,s\beta\,{m_{{{\it KK}}}}^{2}+11\,\beta
\,{m_{{{\it KK}}}}^{4}}{ {m_{{{\it KK}}}}^{2}\pi \,{s}^{2}{\beta}^{2}}}$ \\  
& & $+{\frac {g^4 c_w^2}{576}}\,{\frac {-4\,{m_{{{\it 
KK}}}}^{2}+s}{\beta\,\pi 
\,{s}
^{2}}}     $ \\ 
$W^3_{\mu} B_{\nu} \rightarrow 
h h$ or $ Z_\mu Z_\nu $ & C.2.2 a,b,c& $
\,{\frac {{g}^{4}{Y_{{h}}}^{2}T_3(h)^2{{\it s}_{{w}}}^{2}}{24\pi \,s\beta\,
{c_{{w}}}^{2}}}$ \\
$ W^\pm_{\mu} B_{\nu} \rightarrow z w $ & C.2.2 a,b,c& 
${\frac {1}{192}}\,{\frac {{g}^{4}{{\it s}_{{w}}}^{2}}{s\pi \,\beta\,{
c_{{w}}}^{2}}}$ \\ 
$ W_{3\mu} B_{\nu} \rightarrow w^+ w- $ & C.2.2 a,b,c& 
${\frac {1}{192}}\,{\frac {{g}^{4}{{\it s}_{{w}}}^{2}}{s\pi \,\beta\,{
c_{{w}}}^{2}}}$ \\ 
$ W^\pm_{\mu} B_{\nu} \rightarrow h w $ & C.2.2 a,b,c& 
${\frac {1}{192}}\,{\frac {{g}^{4}{{\it s}_{{w}}}^{2}}{s\pi \,\beta\,{
c_{{w}}}^{2}}}$ \\ 
$W^\pm_{\mu} W^{(3)}_{\nu} \rightarrow h w $ & C.2.2 a,b,c,d& $-{\frac 
{1}{576}}\,{\frac {{g}^{4} \left( -s+4\,{m_{{{\it KK}}}}^{2}
 \right) }{\pi \,{s}^{2}\beta}}$ \\ 
\hline
\end{tabular}
\end{center}

\subsubsection{Fermion Coannihilation}

\begin{center}
\begin{tabular}{|c|c|c|}
\hline
Process & Diagrams & Total cross section \\ 
\hline
$f B_{\nu} \rightarrow f (A_{\nu}+Z_\nu) $& C.4.1 a,c & $  
\,{\frac {g^2 g_1^2 Y^2(T_3^2+\tan^2(\theta_W) Y^2) \left ({ Y_f}^{2} 
\left( 6\,L-\beta
 \right)\right ) }{96\pi \,s{\beta}^{2}}} $ \\
$f W^{(3)}_{\nu} \rightarrow f (A_{\nu}+Z_\nu) $& C.4.1 a,c & $  
\,{\frac {g^4T_3^2(T_3^2+\tan^2(\theta_W) Y^2) \left ({{\it Ye}_{{L}}}^{2} \left( 6\,L-\beta
 \right)\right ) }{96\pi \,s{\beta}^{2}}} $ \\
$f_L^{-} B_{\nu} \rightarrow f_L^{+} W^-_{\nu} $& C.4.1 a,c & $  
\,{\frac {g^2 g_1^2 \left ({{\it Ye}_{{L}}}^{2} \left( 6\,L-\beta
 \right)\right ) }{192\pi \,s{\beta}^{2}}} $ \\
$f_L^{-} W^{3}_{\nu} \rightarrow f_L^{+} W^-_{\nu} $& C.4.1 a,b,c & $ 
\,{\frac {g^4 \left (-26\,{m_{{{\it KK}}}}^{2}L+32\,s\beta+23\,
\beta\,{m_{{{\it KK}}}}^{2}\right )}{768 \pi \,s{\beta}^{2}{m_{{{\it KK}}}}^{2}}} 
$ \\
$W_{\mu}^{-} L_L^+ \rightarrow L_L^{-} (Z_{\mu}+A_{\mu}) $ & C.4.1 a,b,c & $ 
\,{\frac {g^4  \left( \left( -32\,{c_{{w}}}^{2}+6 \right) {m_{{{
\it KK}}}}^{2}L+ \left( 24\,\beta\,{c_{{w}}}^{2}-\beta \right) {m_{{{
\it KK}}}}^{2}+32\,{c_{{w}}}^{2}s\beta\right )}{768 \pi 
\,s{\beta}^{2}{m_{{{\it KK
}}}}^{2}{c_{{w}}}^{2}}}$ \\
$W_{\mu}^{+} L_L^- \rightarrow L_L^{+} (Z_{\mu}+A_{\mu}) $ & C.4.1 a,b,c & as 
above \\
$W_{\mu}^{-} Q_L^+ \rightarrow Q_L^{-} (Z_{\mu}+A_{\mu}) $ & C.4.1 a,b,c & $ 
-{\frac {{g}^{4} \left( -6\,{m_{{{\it KK}}}}^{2}L+
240\,{c_{{w}}}^{2}{m_{{{\it KK}}}}^{2}L+\beta\,{m_{{{\it KK}}}}^{2}-
208\,\beta\,{c_{{w}}}^{2}{m_{{{\it KK}}}}^{2}-288\,{c_{{w}}}^{2}s\beta
 \right) }{6912 {c_{{w}}}^{2}\pi \,s{\beta}^{2}{m_{{{\it KK}}}}^{2}}}$\\
$W_{\mu}^{+} Q_L^- \rightarrow Q_L^{+} (Z_{\mu}+A_{\mu}) $ & C.4.1 a,b,c & as 
above \\
$f_L^{-} W^-_{\nu} \rightarrow f_L^- W^-_{\nu} $& C.4.1 a,b &
 ${\frac {1}{192}}\,{\frac {{g}^{4} \left( 3\,{m_{{{\it KK}}}}^{2}L+4\,
\beta\,s \right) }{\pi \,s{\beta}^{2}{m_{{{\it KK}}}}^{2}}}$ \\
$f_L^{+} W^-_{\nu}, \rightarrow f_L^+ W^-_{\nu} $& C.4.1 b,c & 
$\frac{g^4}{2}{\frac {-16\,L{m_{{{\it KK}}}}^{2}+11\,\beta\,{m_{{{
\it KK}}}}^{2}+8\,s\beta}{192\pi \,s{\beta}^{2}{m_{{{\it KK}}}}^{2}}}
$ \\  
$f_L^{-} W^+_{\nu}, \rightarrow f_L^- W^+_{\nu} $& C.4.1 b,c & as above \\
$W_{3 \mu} Q \rightarrow g_\mu q$ & C.4.1 a,c &
$\frac{g^2T_3^2 g_s^2}{72} \frac{ 6L- \beta}{s \pi \beta^2} $ \\ 
$B_\mu Q \rightarrow g_\mu q$ & C.4.1 a,c &
$\frac{g_1^2Y_Q^2 g_s^2}{72} \frac{ 6L- \beta}{s \pi \beta^2} $ \\ 
$W_\mu^\pm Q \rightarrow g_\mu q'$ & C.4.1 a,c &
$\frac{g^2 g_s^2}{36} \frac{ 6L- \beta}{s \pi \beta^2} $ \\ 
$g_{\mu} q \rightarrow q g_{\mu} $& C.4.1 a,b,c & ${\frac 
{g_s^4 \left(25\,\beta\,{m_{KK}}^{2}-24\,L{m_{KK}}^{
2}+36\,s\beta \right )}{1296\pi \,s{\beta}^{2}{m_{KK}}^{2}}}
$ \\  
$g_{\mu} q \rightarrow q A_{\mu}, Z_{\mu}, W_{\mu} $& C.4.1 a,c &
$\frac{g_s^2 g_{A,Z,W}^2}{2(N_c^2-1)}
{\frac {-\beta+6\,L}{96 \pi \,s{\beta}^{2}}}
$ \\  
\hline
\end{tabular}
\end{center}

\subsection{Higgs Coannihilation}

\begin{center}
\begin{tabular}{|c|c|c|}
\hline
Process & Diagrams & Total cross section \\ 
\hline
$h  W^{3}_{\nu} \rightarrow h  Z_{\nu}  $& C.5.1 a,b,c & $\,{\frac 
{g^4}{96 \pi c_{w}^2 \,s\beta}}
$ \\
$h  B_{\nu} \rightarrow h  Z_{\nu}  $& C.5.1 a,b,c & $\,{\frac 
{g^4 s_w^2}{96 \pi c_{w}^4 \,s\beta}}
$ \\
$z W^{3}_{\nu} \rightarrow z Z_{\nu}  $& C.5.1 a,b,c & $\,{\frac 
{g^4 }{96\pi c_{w}^2 \,s\beta}}
$ \\
$z B_{\nu} \rightarrow z Z_{\nu}  $& C.5.1 a,b,c & $\,{\frac 
{g^4 s_w^2}{96\pi c_{w}^4 \,s\beta}}
$ \\
$\phi (\mbox{or } z) W^{(3)}_{ \nu} \rightarrow w^+ W^-_{\nu}  $& C.5.1 
a,b,c,d & $
\,\frac{g^4}{96}{\frac {-4\,{m_{{{\it KK}}}}^{2}L+\beta\,{m_{{{\it 
KK}}}}^{2}+4
\,s\beta}{\pi \,s{\beta}^{2}{m_{{{\it KK}}}}^{2}}}$ \\
$\phi (\mbox{or } z) B_{ \nu} \rightarrow w^+ W^-_{\nu}  $& C.5.1 a,b,c & $  
\,{\frac {g^4 s_w^2}{96 c_w^2 \beta\,\pi \,s}}
  $ \\ 
$w W^3_{\nu}, B_{\nu} \rightarrow w (Z_{\nu}+ A_\nu)  $& C.5.1 a,b,c & 
$\,{\frac {g^4}{96\pi \,{c_{{w}}
}^{2}s\beta}}$ \\
$w B_{\nu} \rightarrow \phi W_{\nu}  $& C.5.1 a,b,c & $
{\frac {g^4}{96}}\,{\frac {{{\it s}_{{w}}}^{2}}{\pi \,s\beta\,{c
_{{w}}}^{2}}} $ \\
$w W^{(3)}_{\nu} \rightarrow \phi W_{\nu}  $& C.5.1 a,b,d & $
{\frac {g^4}{96}}\,{\frac { -4\,L{m_{{{\it KK}}}}^{2}+
\beta\,{m_{{{\it KK}}}}^{2}+4\,s\beta }{\pi \,s{\beta}^{2}{m_{
{{\it KK}}}}^{2}}} $ \\
$\phi(\mbox{or }z) W_{\nu} \rightarrow \phi(\mbox{or }z) W_{\nu}  $& 
C.5.1 a,b,c & 
$\,{\frac {g^4}{96 \pi \,s\beta}}
$ \\
$z (\mbox{or }h) W_{\nu} \rightarrow h (\mbox{or }z) W_{\nu}  $& C.5.1 
a,b,d & 
$\,{\frac {g^4 \left(4\,s\beta+\beta\,{m_{{{\it KK}}}}^{2}-4\,L{m_{{{\it 
KK}}
}}^{2}\right)}{96 s\pi \,{\beta}^{2}{m_{{{\it KK}}}}^{2}}}
$ \\ 
$w^- W^+_{\nu} \rightarrow z (\mbox{or } h ) (Z_{\nu}+A_\nu)  $& C.5.1 
a,b,c,d & 
$ {\frac {g^4}{96}}\,{\frac { \left( -4\,L{c_w}^{2}+\beta \right) 
{m_{{{\it 
KK}}}}^{2}+4\,{c_w}^{2}s\beta}{\pi \,s{\beta}^{2}{c_w}^{2}{m_{{{\it KK}}}}
^{2}}}$ \\
$w^- W^+_{\nu} \rightarrow w^- W^+_{\nu}  $& C.5.1 b,c,d &  $\,{\frac 
{g^4}{24\pi \,\beta\,{m_{{{\it KK}}}}^{2}}}$ \\
$w^+ W^+_{\nu} \rightarrow w^+ W^+_{\nu}  $& C.5.1 a,c,d &  $ 
\frac{g^4}{24}{\frac { \left( -2\,L+\beta \right) {m_{{{\it 
KK}}}}^{2}+s\beta}
{ \pi \,s{\beta}^{2}{m_{{{\it KK}}}}^{2}}}$ \\
$f \phi \rightarrow f \phi $& C.6.1 a & $ \frac{(g_{f} g_{ 
\phi})^2}{8}\,{\frac {s\beta-L{m_{{{\it 
KK}}}}^{2}}{\pi \,s{\beta}^{2}{m_{{{
\it KK}}}}^{2}}}
$ \\  
\hline
\end{tabular}
\end{center}

In the last line, $g_f$ and $g_{\phi}$ denote the couplings of 
the appropriate $t$-channel gauge boson to the fermions and the scalars, 
respectively.


\end{document}